\documentclass[prd,twocolumn,floatfix,amsmath,nofootinbib,amssymb,superscriptaddress]{revtex4}
\usepackage{graphicx,color,dcolumn,booktabs,bm}
\usepackage{longtable,lscape}
\usepackage{pdfpages}
\usepackage{txfonts}
\usepackage{overpic}
\usepackage{amssymb}
\usepackage{makecell}
\usepackage{indentfirst}
\usepackage{feynmf}
\usepackage{slashed}
\usepackage[utf8]{inputenc}
\usepackage{cases}
\usepackage{color}
\usepackage{multirow}
\usepackage{threeparttable}
\usepackage{epstopdf}
\usepackage{enumerate}
\usepackage{subfigure}
\usepackage{diagbox}
\usepackage{graphicx,color,dcolumn,booktabs,bm}
\usepackage{mathrsfs}
\usepackage{cancel}
\usepackage{float}
\usepackage[misc]{ifsym}
\usepackage[colorlinks,
citecolor=blue,
anchorcolor=red,
menucolor=red,
linkcolor=red,
filecolor=red,
runcolor=red,
urlcolor=blue,
frenchlinks=red]{hyperref}
\usepackage{array}
\usepackage{booktabs}
\usepackage{amsmath}
\usepackage{slashed}

\begin{document}
\title{Form factors and weak decays of $\Lambda_b\to\Lambda_c^*(2595)$ and $\Xi_b\to\Xi_c^*(2790)$}
\author{Jie Lu}
\email{l17693567997@163.com}
\affiliation{School of Physics, Southeast University, Nanjing 210094, People's Republic of China}

\author{Dian-Yong Chen}
\email{chendy@seu.edu.cn}
\affiliation{School of Physics, Southeast University, Nanjing 210094, People's Republic of China}
\affiliation{Lanzhou Center for Theoretical Physics, Lanzhou University, Lanzhou 730000, People's Republic of China}

\author{Guo-Liang Yu}
\email{yuguoliang2011@163.com}
\affiliation{Department of Mathematics and Physics, North China Electric Power University, Baoding 071003, People's Republic of China}
\affiliation{Hebei Key Laboratory of Physics and Energy Technology, North China Electric Power University, Baoding 071000, China}

\author{Zhi-Gang Wang}
\email{zgwang@aliyun.com}
\affiliation{Department of Mathematics and Physics, North China Electric Power University, Baoding 071003, People's Republic of China}
\affiliation{Hebei Key Laboratory of Physics and Energy Technology, North China Electric Power University, Baoding 071000, China}
\date{\today }

\begin{abstract}
In this article, we reanalyze the mass and pole residue of $\Lambda$-type singly heavy baryons with spin-parity $J^P=1/2^{\pm}$ using two-point QCD sum rules. Then, we investigate the electroweak transition form factors of $\Lambda_b\to\Lambda^*_c(2595)$ and $\Xi_b\to\Xi^*_c(2790)$ within the framework of three-point QCD sum rules. At the hadron side, all possible couplings of the interpolating currents to hadronic states are taken into account, while at the QCD side, both the perturbative part and vacuum condensate contributions up to dimension-8 are included. Using the obtained form factors, we calculate the decay widths and branching ratios for the semileptonic decays $\Lambda_b\to\Lambda^*_c(2595)\ell\bar{\nu}_\ell$ and $\Xi_b\to\Xi^*_c(2790)\ell\bar{\nu}_\ell$ ($\ell=e,\mu$ and $\tau$), as well as for the corresponding nonleptonic decays $\Lambda_b\to\Lambda^*_c(2595)\pi[\rho]$, $\Lambda_b\to\Lambda^*_c(2595)K[K^*]$, $\Xi_b\to\Xi^*_c(2790)\pi[\rho]$ and $\Xi_b\to\Xi^*_c(2790)K[K^*]$. Our estimates for the branching ratios of $\Lambda_b\to\Lambda^*_c(2595)\ell\bar{\nu}_\ell$ are comparable with the experimental data and other theoretical results. These results can provide valuable references for future measurements. In addition, we present predictions for several observables in these semileptonic decays, including the lepton universality ratio, the forward-backward asymmetry $A_{FB}$, and the asymmetry parameter $\alpha$, which can provide useful references for future experiments to test the Standard Model and probe new physics beyond it.
\end{abstract}


\maketitle

\section{Introduction}\label{sec1}

Form factors play a fundamental role in understanding weak decays of hadrons. They are essential not only for determining decay widths but also for revealing the underlying hadronic dynamics. Within the Standard Model, the Cabibbo-Kobayashi-Maskawa (CKM) matrix element $V_{cb}$ remains imprecisely determined. Traditionally, its value has been extracted mainly from exclusive and inclusive semileptonic decays of bottom mesons to charmed mesons~\cite{Gambino:2011fz,Ricciardi:2019zph,Fael:2024rys,ParticleDataGroup:2026mpi}. To achieve a complementary and more robust determination, data from semileptonic decays of bottom baryons to charmed baryons must be included. Currently, the only experimental data on such decays come from measurements of the $\Lambda_b$ baryon decaying to the ground state $\Lambda_c$ and its excited states~\cite{ParticleDataGroup:2026mpi}. Although various models have explored $\Lambda_b$ decays to the ground-state $\Lambda_c$ with $J^P=1/2^+$, excited $\Lambda_c$ states, such as $\Lambda_c^*(2595)$ and $\Lambda_c^*(2625)$ with $J^P=1/2^-$ and $3/2^-$, respectively, also contribute significantly to the semileptonic decay of $\Lambda_b$. Investigating semileptonic decays of bottom baryons with excited $\Lambda_c$ states in the final state will further enrich our understanding of these processes and provide valuable insights into bottom-baryon decay dynamics. The corresponding nonleptonic decays, which involve both factorizable and non-factorizable contributions, are equally important and should be analyzed as well. Moreover, the weak decays of the $\Xi_b$ baryon are equally significant. While $\Xi_b$ shares similar properties with $\Lambda_b$, the presence of an additional strange quark provides a complementary probe of SU(3) flavor symmetry breaking and heavy-quark dynamics. Although experimental data remain scarce, future observations are expected to be highly promising.

Experimentally, the CDF Collaboration reported a measurement of the ratio of branching fractions $\mathcal{B}[\Lambda_b^0 \to \Lambda_c^+ \mu^- \bar{\nu}_\mu]/\mathcal{B}[\Lambda_b^0 \to \Lambda_c^+ \pi^-]$ in 2009. In the same report, they also presented the branching fractions for the semileptonic decays $\Lambda_b^0 \to \Lambda^*_c(2595)^+ \mu^- \bar{\nu}_\mu$ and $\Lambda_b^0 \to \Lambda^*_c(2625)^+ \mu^- \bar{\nu}_\mu$, relative to that of $\Lambda_b^0 \to \Lambda_c^+ \mu^- \bar{\nu}_\mu$~\cite{CDF:2008hqh}. To date, the Particle Data Group (PDG) lists the absolute branching fractions for these two semileptonic channels as $\mathcal{B}[\Lambda_b^0 \to \Lambda_c^*(2595)^+ \ell^- \bar{\nu}_\ell] = (7.9^{+4.0}_{-3.5}) \times 10^{-3}$ and $\mathcal{B}[\Lambda_b^0 \to \Lambda_c^*(2625)^+ \ell^- \bar{\nu}_\ell] = (1.3^{+0.6}_{-0.5}) \times 10^{-2}$~\cite{ParticleDataGroup:2026mpi}. In addition, the PDG quotes the product of the branching ratios for the decay $\Lambda_b^0 \to \Lambda_c^{*+} \pi^-$ and the subsequent decay $\Lambda^{*+}_c \to \Lambda_c^+ \pi^+ \pi^-$ as $(3.4\pm1.4)\times10^{-4}$ for $\Lambda^*_c(2595)$ and $(3.3\pm1.3)\times10^{-4}$ for $\Lambda^*_c(2625)$, respectively~\cite{ParticleDataGroup:2026mpi}. For the $\Xi_b$ baryon, in contrast, no experimental results are yet available for weak decays with a ground or excited $\Xi_c$ baryon in the final state.

\begin{figure}
	\centering
	\includegraphics[width=9cm]{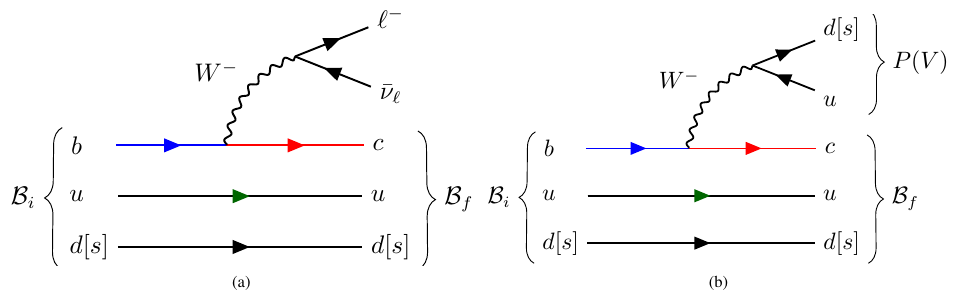}
	\caption{Feynman diagrams for (a) the semileptonic decay $\Lambda_b[\Xi_b] \to \Lambda_c^*[\Xi_c^*] \ell^- \bar{\nu}_\ell$ and (b) the $W$-boson emission diagram ($T$) for the nonleptonic decay $\Lambda_b[\Xi_b] \to \Lambda_c^*[\Xi_c^*] P[V]$.}\label{Weakdecay}
\end{figure}

On the theoretical side, the weak decays of $\Lambda_b[\Xi_b]$ baryons to ground and excited $\Lambda_c[\Xi_c]$ states, including both semileptonic and nonleptonic processes, are interesting and valuable. In the case of semileptonic decays, the $b$ quark decays via $b \to c\ell\bar{\nu}_\ell$ ($\ell=e,\mu,\tau$), and the final $c$ quark combines with the spectator diquark to form the final baryon, as shown in Fig.~\ref{Weakdecay}(a). For nonleptonic decays, the final-state mesons involved are only light pseudoscalar and vector mesons, and the situation is more complex since both factorizable and non-factorizable contributions are present. At tree level, the factorizable contribution is dominated by the $W$-boson emission ($T$) topology, as shown in Fig.~\ref{Weakdecay}(b). In bottom baryon decays, the remaining non-factorizable long-distance contributions, including the color-suppressed ($C^{(')}$), exchange ($E$), and baryon-exchange ($B$) topologies, are suppressed by $\Lambda_{\mathrm{QCD}}/m_b$ according to soft-collinear effective theory~\cite{Mantry:2003uz,Leibovich:2003tw}. Therefore, we focus only on the $T$-diagram contribution. Nevertheless, since both semileptonic and nonleptonic decays of bottom baryons involve strong and weak interactions, they cannot be treated purely perturbatively. Over the years, various non-perturbative approaches have been developed to study these processes, including lattice QCD (LQCD)~\cite{Detmold:2015aaa,Meinel:2021rbm}, QCD sum rules (QCDSR)~\cite{Azizi:2018axf,Zhao:2020mod,Neishabouri:2025abl,Lu:2025gol,Lu:2026qkk}, light-cone QCD sum rules (LCSR)~\cite{Miao:2022bga,Duan:2022uzm,Duan:2024lnw,Duan:2026smz}, perturbative QCD (pQCD)~\cite{Shih:1999yh,Rui:2025iwa,Li:2025rsm}, various quark models~\cite{Ebert:2006rp,Gutsche:2014zna,Faustov:2016pal,Faustov:2018ahb,Zhao:2018zcb,Gutsche:2018nks,Li:2021qod,Lih:2026uhh}, and others~\cite{Dutta:2015ueb,Patel:2025rhg,Zhang:2022bvl}.

As one of the most powerful non-perturbative approaches for dealing with hadronic parameters, QCDSR are time-honored and widely used in studying hadron properties~\cite{Shifman:1978bx,Shifman:1978by,Reinders:1984sr,Colangelo:2000dp,Wang:2025sic}. In particular, three-point QCDSR and LCSR are both extensively employed to extract form factors and are complementary to each other. Three-point QCDSR are particularly suited to baryon-to-baryon transitions, as they treat the transition matrix elements systematically through a local operator product expansion (OPE), including perturbative and vacuum condensate contributions, while the Borel transformation suppresses the contributions from higher resonances and continuum states, enabling reliable form-factor extraction in the space-like region ($Q^2=-q^2>0$) and subsequent extrapolation to the time-like region ($q^2>0$)~\cite{Reinders:1984sr,Colangelo:2000dp}. However, their accuracy depends on the Borel and continuum threshold parameters. In addition, the truncation of the OPE and model-dependent extrapolations such as the $z$-series expansion also introduce uncertainties. The LCSR, by contrast, are powerful for meson-to-meson transitions at large recoil (corresponding to small $q^2$), since they encode the long-distance dynamics in meson light-cone distribution amplitudes and are less sensitive to the Borel parameters~\cite{Colangelo:2000dp,Braun:1997kw,Khodjamirian:2023wol}. Their application to baryon-to-baryon transitions is limited, because baryon light-cone distribution amplitudes are poorly known and much more complex than their mesonic counterparts. Consequently, for baryon-to-baryon weak decays, three-point QCDSR remain the method of choice, whereas LCSR are more suitable for meson-to-meson processes. 

In our previous work, the form factors and corresponding weak decays of singly and doubly heavy baryons are analyzed by three-point QCDSR~\cite{Lu:2025gol,Yu:2026tbk,Lu:2026qkk,Yu:2026wsl,Yang:2026vax}, where all possible coupling of interpolating current to hadronic states are
considered at hadron side and the Dirac structures independent form factors are obtained. As continuation and expansion of these works, we analyze the vector and axial vector form factors of $\Lambda_b\to\Lambda^*_c(2595)$ and $\Xi_b\to\Xi^*_c(2790)$ by using three-point QCDSR in the present work. With obtained form factors, the corresponding semileptonic and nonleptonic decay processes are also analyzed.

This paper is organized as follows. After introduction in Sec.~\ref{sec1}, we firstly analyze the mass and pole residue of positive and negative $\Lambda$-type singly heavy baryons with two-point QCDSR in Sec.~\ref{sec2}. Then, we analyze the electroweak transition from factors for $\Lambda_b\to\Lambda^*_c(2595)$ and $\Xi_b\to\Xi_c^{*}(2790)$ with the three-point QCDSR in Sec.~\ref{sec3}. With the estimated form factors, we analyze the corresponding semileptonic and nonleptonic decay processes in Sec.~\ref{sec4}. Sec.~\ref{sec5} is devoted to present the numerical results and discussions and Sec.~\ref{sec6} is the conclusion part.

\section{Two-point QCD sum rules for mass and pole residue of $\Lambda$-type single heavy baryon}\label{sec2}
In this section, we improve our previous calculations of the masses and pole residues for $\Lambda$-type singly heavy baryons with spin-parity $1/2^\pm$ using the two-point QCDSR.

The interpolating current of singly heavy baryons can be uniformly expressed as following form:
\begin{eqnarray}
	J(x)=\varepsilon_{ijk}q_{1}^{iT}(x)\mathcal{C}\Gamma_1q_2^j(x)\Gamma_2Q^{k}(x),
\end{eqnarray}
where $\varepsilon_{ijk}$ is the 3-dimension Levi-Civta tensor, $i$, $j$ and $k$ are color indices, $q_i$ and $Q$ denote the light $u$, $d$ or $s$ and heavy $c$ or $b$ quarks, respectively. $\mathcal{C}$ is the charge conjugation operator, $\Gamma_1$ and $\Gamma_2$ are different Dirac matrices ($I$, $\gamma_5$, $\gamma_\mu$, $\gamma_\mu\gamma_5$ and $\sigma_{\mu\nu}(\gamma_5)$) combined with the covariant derivative operator, used to describe the orbital spin information of light diquark systems as well as between light diquark systems and heavy quarks. For the $\Lambda_Q$ and $\Xi_Q$ baryons with $J^P=1/2^+$, we commonly take $\Gamma_1=\gamma_5$ and $\Gamma_2=I$.

If a baryon interpolating current $J(x)$ has the spin-parity $J^P=1/2^+$, then $\gamma_5J(x)$ will have the spin-parity $J^P=1/2^-$, because multiplying $\gamma_5$ can change the parity of the current $J(x)$. Effectively separating the contributions of positive- and negative-parity states is an important challenge in the QCDSR study of baryons. In 1993, Bagan et al. used the infinite heavy-quark mass limit $m_Q\to\infty$ to achieve an unambiguous separation of the positive- and negative-parity heavy baryon contributions~\cite{Bagan:1993ii}. In 1996, Jido et al. proposed a new scheme to separate the contributions of positive- and negative-parity light-flavor baryons~\cite{Jido:1996ia}. Following the Ref.~\cite{Jido:1996ia}, we firstly write down the two-point correlation functions $\Pi_\pm$,
\begin{eqnarray}\label{eq:2}
	\Pi_\pm(p)=i\int d^4x e^{ip\cdot x}\langle 0|\mathcal{T}[J_\pm(x)\bar{J}_\pm(0)]|0\rangle,
\end{eqnarray}
where $\mathcal{T}$ denote time order operator, $\bar{J}=J^\dagger\gamma_0$, the subscripts $\pm$ is introduced to represent the positive and negative parity, respectively, $J_-=\gamma_5J_+$. The correlation functions can be decomposed as:
\begin{eqnarray}
 	\Pi_\pm(p)=\Pi^1(p^2)\slashed{p}\pm\Pi^2(p^2),
\end{eqnarray}
the current $J_+$ can couple to both positive- and negative-parity baryons, the coupling can be expressed as:
\begin{eqnarray}\label{eq:4}
	\notag
	\langle 0|J_+|B^+\rangle&=&\lambda_+u_+,\\
	\langle 0|J_+|B^-\rangle&=&\lambda_-\gamma_5u_-,
\end{eqnarray}
where $\lambda_\pm$ and $u_\pm$ are pole residue and spinor wave function, $B^\pm$ denote the positive- and negative-parity baryons, respectively. Similarly, current $J_-$ can also couple to both positive- and negative-parity baryons as:
\begin{eqnarray}\label{eq:5}
	\notag
	\langle 0|J_-|B^-\rangle&=&\lambda_-u_-,\\
	\langle 0|J_-|B^+\rangle&=&\lambda_+\gamma_5u_+,
\end{eqnarray}
According to the Eqs.~(\ref{eq:4}) and (\ref{eq:5}), we can obtain the correlation function at hadron side. Taking the $\Pi_+$ as an example, the correlation function can be written as:
\begin{eqnarray}
	\Pi_+(p)=\lambda_+^2\frac{\slashed{p}+m_+}{m_+^2-p^2}+\lambda_-^2\frac{\slashed{p}-m_-}{m_-^2-p^2}+h.c.
\end{eqnarray}
where $m_\pm$ are the baryon masses. $h.c.$ denote the contributions from higher resonances and continuum states of baryons.

If we discuss the correlation function in the rest frame of the baryon $\vec{p}=0$, we have
\begin{eqnarray}
	\notag
	\frac{1}{\pi}\lim_{\epsilon\to0}\Pi_+(p_0+i\epsilon)&=&\lambda_+^2\frac{\gamma_0+1}{2}\delta(p_0-m_+)\\
	\notag
	&&+\lambda_-^2\frac{\gamma_0-1}{2}\delta(p_0-m_-)+h.c.\\
	&=&\gamma_0A(p_0)+B(p_0)+h.c.,
\end{eqnarray}
with
\begin{eqnarray}
	\notag
	A(p_0)&&=\frac{\lambda_+^2\delta(p_0-m_+)+\lambda_-^2\delta(p_0-m_-)}{2},\\
	B(p_0)&&=\frac{\lambda_+^2\delta(p_0-m_+)-\lambda_-^2\delta(p_0-m_-)}{2}.
\end{eqnarray}
From above equation, one can find that the contribution $A(p_0)+B(p_0)$ ($A(p_0)-B(p_0)$) contains contributions from the positive- (negative-) parity states only. Then, the OPE will carried out at large $P^2\to-p_0^2$ region. By using the dispersion relation, the QCD spectral densities $\rho^A(p_0)$ and $\rho^B(p_0)$ at QCD side can be obtained. Finally, a clean QCD sum rules for positive parity baryon can be obtained as the following forms by using Borel transform and quark-hadron duality,
\begin{eqnarray}
	\notag
	&&\int_{\sqrt{s_{\min}}}^{\sqrt{s_0}}dp_0\left[A(p_0)+B(p_0)\right]\exp\left(-\frac{p_0^2}{T^2}\right)\\
	&&=\int_{\sqrt{s_{\min}}}^{\sqrt{s_0}}dp_0\left[\rho^A(p_0)+\rho^B(p_0)\right]\exp\left(-\frac{p_0^2}{T^2}\right),
\end{eqnarray}
where $\sqrt{s_{\min}}$ and $\sqrt{s_0}$ are the kinematic and continuum threshold, respectively, $T^2$ is the Borel parameter. This method was successfully used to analyze the mass difference of positive- and negative-parity nucleons. Subsequently, we extended it to study singly, doubly, and triply heavy baryons with spin-parity $1/2^\pm$ and $3/2^\pm$~\cite{Wang:2009cr,Wang:2010hs,Wang:2010fq,Wang:2010vn,Wang:2010it,Wang:2011ae}. 

The procedure mentioned in Ref.~\cite{Jido:1996ia} are semi-analytical, and moreover, by taking the baryon three-momentum as zero, this method loses Lorentz covariance. In previous work, we found that for both positive- and negative-parity baryons, the Borel platforms obtained using this method were not particularly flat. In addition, the continuum threshold parameter needed to be set relatively high in order to reproduce the experimental masses of these baryons. In Ref.~\cite{Wang:2015epa}, we introduced an analytical procedure to address this problem. For correlation function $\Pi_+$, the spectral densities at hadron side can be obtained through the following dispersion relation,
\begin{eqnarray}
	\notag
	\frac{1}{\pi}\lim_{\epsilon\to0}\Pi_+(s+i\epsilon)&&=\left[\lambda_+\delta(s-m_+^2)+\lambda_-\delta(s-m_-^2)\right]\slashed{p}\\
	\notag
	&&+\left[m_+\lambda_+\delta(s-m_+^2)-m_-\lambda_-\delta(s-m_-^2)\right]+h.c.\\
	&&=\rho_H^1(s)\slashed{p}+\rho_H^2(s),
\end{eqnarray}
here, the subscript $H$ denotes the hadron side. By introducing the weighting factor $\sqrt{s}$ and doing the Borel transform, the QCD sum rules at hadron side can be obtained as:
\begin{eqnarray}
	\notag
	&&\int_{s_{\min}}^{s_0}ds\left[\sqrt{s}\rho_H^1(s)\pm\rho_H^2(s)\right]\exp\left(-\frac{s}{T^2}\right)\\
	&&=2\lambda_\pm^2m_\pm\exp\left(-\frac{m_\pm^2}{T^2}\right),
\end{eqnarray}
In Ref.~\cite{Wang:2020mxk}, we use this approach to analyze the $1S$ and $2S$ $\Lambda$-type singly heavy baryons $\Lambda_Q$ and $\Xi_Q$ with spin-parity $1/2^+$. A more reasonable continuum threshold and a flatter Borel platform have been obtained. In the present work, we continue to use this method to reanalyze $\Lambda$-type singly baryons with positive and negative parity, and increase the dimension of the OPE from 10 to 15 dimensions.

Firstly, the positive-parity $\Lambda$-type singly heavy baryons can be interpolated by the following current,
\begin{eqnarray}\label{eq:12}
	J_{\Lambda_Q[\Xi_Q]}(x)=\varepsilon_{ijk}u^{iT}(x)\mathcal{C}\gamma_5d[s]^j(x)Q^{k}(x).
\end{eqnarray}
We introduce the interpolating current into the two-point correlation function in Eq.~(\ref{eq:2}) and insert the hadronic complete sets coupled to it. After performing the integrals in coordinate space, isolating the contributions of the ground states from excited states and using the dispersion relation, the two-point correlation function at hadron side can be written as,
\begin{eqnarray}\label{eq:13}
	\notag
	\Pi^H_+(p)&&=\lambda_{\Lambda_Q[\Xi_Q]}^2\frac{\slashed{p}+m_{\Lambda_Q[\Xi_Q]}}{m_{\Lambda_Q[\Xi_Q]}^2-p^2}+\lambda_{\Lambda^*_Q[\Xi^*_Q]}^2\frac{\slashed{p}-m_{\Lambda^*_Q[\Xi^*_Q]}}{m_{\Lambda^*_Q[\Xi^*_Q]}^2-p^2}+h.c.\\
	\notag
	&&=\left(\frac{\lambda_{\Lambda_Q[\Xi_Q]}^2}{m_{\Lambda_Q[\Xi_Q]}^2-p^2}+\frac{\lambda_{\Lambda^*_Q[\Xi^*_Q]}^2}{m_{\Lambda^*_Q[\Xi^*_Q]}^2-p^2} \right)\slashed{p}\\
	&&+\left(\frac{\lambda_{\Lambda_Q[\Xi_Q]}^2m_{\Lambda_Q[\Xi_Q]}}{m_{\Lambda_Q[\Xi_Q]}^2-p^2}-\frac{\lambda_{\Lambda^*_Q[\Xi^*_Q]}^2m_{\Lambda^*_Q[\Xi^*_Q]}}{m_{\Lambda^*_Q[\Xi^*_Q]}^2-p^2} \right)+h.c.
\end{eqnarray}
where the $\Lambda_Q[\Xi_Q]$ and $\Lambda^*_Q[\Xi^*_Q]$ denote the positive- and negative-parity $\Lambda$-type singly heavy baryon, respectively.

At QCD side, we contract the quark fields with Wick’s theorem, and express the correlation function in terms of the full quark propagators. After these steps, the correlation function at QCD side can be written as,
\begin{eqnarray}\label{eq:14}
	\notag
	\Pi^{\rm{QCD}}_+(p)&=&i\varepsilon_{ijk}\varepsilon_{i'j'k'}\int d^4xe^{ip\cdot x}\\
	&&\times\mathrm{Tr}\{\gamma_5\mathcal{C}U^{ik'T}(x)\mathcal{C}\gamma_5D[S]^{jj'}(x)\}Q^{ki'}(x),
\end{eqnarray}
where $U[D]^{ij}(x)$, $S^{ij}(x)$ and $Q^{ij}(x)$ are the full propagators of $u(d)$, $s$ and $c(b)$ quarks which have the following forms~\cite{Reinders:1984sr},
\begin{eqnarray}\label{eq:15}
	\notag
	U[D]^{ij}(x) &&= \frac{i\delta ^{ij}\slashed x}{2\pi ^2x^4} - \frac{\delta ^{ij}\left\langle \bar qq \right\rangle}{12} - \frac{\delta ^{ij}x^2\left\langle \bar qg_s\sigma Gq \right\rangle }{192}\\
	\notag
	&&- \frac{ig_sG_{\alpha \beta }^at_{ij}^a(\slashed x\sigma ^{\alpha \beta } + \sigma ^{\alpha \beta }\slashed x)}{32\pi ^2x^2} - \frac{\left\langle \bar q^j\sigma ^{\mu \nu }q^i \right\rangle \sigma _{\mu \nu }}{8} \\
	\notag
	&&-\frac{\delta ^{ij}x^4\left\langle \bar qq \right\rangle\left\langle g_s^2GG \right\rangle}{27648}+...,
\end{eqnarray}
\begin{eqnarray}
	\notag
	S^{ij}(x) &&= \frac{i\delta ^{ij}\slashed x}{2\pi ^2x^4} - \frac{\delta ^{ij}m_s}{4\pi ^2x^2} - \frac{\delta ^{ij}\left\langle \bar ss \right\rangle }{12} + \frac{i\delta ^{ij}\slashed xm_s\left\langle \bar ss \right\rangle }{48}\\
	\notag
	&&- \frac{\delta ^{ij}x^2\left\langle \bar sg_s\sigma Gs \right\rangle }{192} + \frac{i\delta ^{ij}x^2\slashed xm_s\left\langle \bar s{g_s}\sigma Gs \right\rangle }{1152}\\
	\notag
	&&- \frac{ig_sG_{\alpha \beta }^at_{ij}^a(\slashed x\sigma ^{\alpha \beta} + \sigma ^{\alpha \beta }\slashed x)}{32\pi ^2x^2} - \frac{\left\langle {\bar s^j\sigma ^{\mu \nu }s^i} \right\rangle \sigma _{\mu \nu }}{8} \\
	\notag
	&&-\frac{\delta ^{ij}x^4\left\langle \bar ss \right\rangle\left\langle g_s^2GG \right\rangle}{27648}+...,
\end{eqnarray}
\begin{eqnarray}
	\notag
	Q^{ij}(x) &&= \frac{i}{(2\pi )^4}\int d^4k e^{ - ik\cdot x} \left\{ \frac{\delta ^{ij}}{\slashed k - m_Q} \right.\\
	\notag
	&&- \frac{g_sG_{\alpha \beta }^nt_{ij}^n}{4}\frac{\sigma ^{\alpha \beta }(\slashed k + m_Q) + (\slashed k + m_Q)\sigma ^{\alpha \beta }}{(k^2 - m_Q^2)^2}\\
	\notag
	&&- \frac{g_s^2(t^at^b)_{ij}G_{\alpha \beta }^aG_{\mu \nu }^b(f^{\alpha \beta \mu \nu } + f^{\alpha \mu \beta \nu} + f^{\alpha \mu \nu \beta })}{4(k^2 - m_Q^2)^5}\\
	&&\left. { + ...} \right\}.
\end{eqnarray}
Here, $\langle\bar{q}(\bar{s})g_s\sigma Gq(s)\rangle=\langle\bar{q}(\bar{s})g_s\sigma_{\mu\nu} G^{\alpha}_{\mu\nu}t^{\alpha}q(s)\rangle$, $t^\alpha=\frac{\lambda^\alpha}{2}$, $\lambda^\alpha$ denote the Gell-Mann matrix and $f^{\alpha \beta \mu \nu }$ can be expressed as:
\begin{eqnarray}\label{eq:16}
	\notag
	f^{\alpha \beta \mu \nu } &&= (\slashed k + m_Q)\gamma ^\alpha (\slashed k + m_Q)\gamma ^\beta(\slashed k + m_Q)\gamma ^\mu(\slashed k + m_Q)\gamma ^\nu\\
	&&\times(\slashed k + m_Q).
\end{eqnarray}
According to the dispersion relation, the correlation functions at QCD side can be expressed as the following form,
\begin{eqnarray}\label{eq:17}
	\notag
	\Pi^{\rm{QCD}}_+(p)&&=\Pi^{1\rm{QCD}}_+(p^2)\slashed{p}+\Pi^{2\rm{QCD}}_+(p^2)\\
	&&=\int_{s_{\min}}^{\infty}\frac{\rho^{1\rm{QCD}}_+(s)}{s-p^2}\slashed{p}+\int_{s_{\min}}^{\infty}\frac{\rho^{2\rm{QCD}}_+(s)}{s-p^2},
\end{eqnarray}
where $s=p^2$, $\rho^{i\rm{QCD}}_+(s)$ ($i=1,2$) denote the QCD spectral which can be expressed as,
\begin{eqnarray}\label{eq:18}
	\rho^{i\mathrm{QCD}}_+(s)&=&\frac{1}{2\pi i}\mathrm{Dis}_s\Pi^{i\mathrm{QCD}}_+(s),
\end{eqnarray}
here, $\rm{Dis}_s$ represents the discontinuity of the variable $s$, and can be written as:
\begin{eqnarray}
	\notag
	\mathrm{Dis}_sf(s)=\lim_{\epsilon\to0}[f(s+i\epsilon)-f(s-i\epsilon)]=2i\mathrm{Im}_{\epsilon\to0}f(s+i\epsilon).
\end{eqnarray}
By matching the Eqs.~(\ref{eq:13}) and (\ref{eq:17}) with quark-hadron duality condition, and then using the scheme in Ref.~\cite{Wang:2020mxk} to separate the contributions of positive- and negative-parity baryons, and performing the Borel transformation for the variable $P^2=-p^2$, we obtain the following two sum rules:
\begin{eqnarray}\label{eq:19}
	\notag
	&&2\lambda_{\Lambda_Q[\Xi_Q]}^2m_{\Lambda_Q[\Xi_Q]}\exp\left(-\frac{m_{\Lambda_Q[\Xi_Q]}^2}{T^2}\right)\\
	&&=\int_{s_{\min}}^{s_0}ds\left[\sqrt{s}\rho^{\rm{1QCD}}_+(s)+\rho^{2\rm{QCD}}_+(s)\right]\exp\left(-\frac{s}{T^2}\right),
\end{eqnarray}
\begin{eqnarray}\label{eq:20}
	\notag
	&&2\lambda_{\Lambda^*_Q[\Xi^*_Q]}^2m_{\Lambda^*_Q[\Xi^*_Q]}\exp\left(-\frac{m_{\Lambda^*_Q[\Xi^*_Q]}^2}{T^2}\right)\\
	&&=\int_{s_{\min}}^{s_0}ds\left[\sqrt{s}\rho^{1\rm{QCD}}_+(s)-\rho^{\rm{2QCD}}_+(s)\right]\exp\left(-\frac{s}{T^2}\right),
\end{eqnarray}
where $T^2$ is the Borel parameter. $s_{\min}$ denotes the kinematic threshold of the baryon and is usually taken as the square of the sum of its constituent quark masses. $s_0$ represents the continuum threshold of the corresponding baryon and usually satisfies the relation $s_0=(m_{\mathcal{B}} + \delta)^2$, where $m_{\mathcal{B}}$ is the mass of the specific baryon state under consideration and $\delta$ is the energy gap between this state and the first excited state in the same channel, typically taking values in the range $0.4-0.8$ GeV. In the present work, we truncated the OPE dimension to 15. Taking the partial derivative of both sides of Eqs.~(\ref{eq:19}) and (\ref{eq:20}) with respect to the variable $-1/T^2$ respectively yields the QCD sum rule equations for mass and pole residues. Here, we only present the sum rule equations for the masses and pole residues of positive-parity baryons, which can be expressed as follows:
\begin{eqnarray}\label{eq:21}
	\notag
	m_{\Lambda_Q[\Xi_Q]}^2 &&= \frac{\int_{s_{\min}}^{s_0} dse^{- s/T^2} s\left[\sqrt{s}\rho^{1\mathrm{QCD}}_+(s) + \rho^{2\mathrm{QCD}}_+(s)\right]}{\int_{s_{\min}}^{s_0} dse^{- s/T^2}\left[\sqrt{s}\rho^{1\mathrm{QCD}}_+(s) + \rho^{2\mathrm{QCD}}_+(s)\right]},\\
	\notag
	\lambda _{\Lambda_Q[\Xi_Q]}^2 &&= \frac{\int_{s_{\min}}^{s_0} dse^ {-s/T^2} \left[\sqrt{s}\rho^{1\mathrm{QCD}}_+(s)+\rho^{2\mathrm{QCD}}_+(s)\right]}{2m_{\Lambda_Q[\Xi_Q]}e^{-m_{\Lambda_Q[\Xi_Q]}^2/T^2}}\\
\end{eqnarray}

In practical calculations, we find that for positive-parity baryons the sum rule in Eq.~(\ref{eq:19}) yields a flat Borel platform and reproduces the experimental mass values. However, for negative-parity baryons, the sum rule in Eq.~(\ref{eq:20}) fails to produce a satisfactory platform. Therefore, a self-consistent scheme is required for the negative-parity case. In Eq.~(\ref{eq:19}), when the continuum threshold $s_0$ for $\Lambda_Q[\Xi_Q]$ is adopted, part of the contribution from the negative-parity pole and the continuum is already included in the spectral density; introducing the weighting factor $\sqrt{s}$ can effectively reduce this contamination. In contrast, when the continuum threshold $s_0$ for $\Lambda^*_Q[\Xi^*_Q]$ is taken in Eq.~(\ref{eq:20}), the positive-parity pole, the continuum, and even lower-lying excited states are all contained in the spectral density, and the weighting factor $\sqrt{s}$ can no longer suppress them effectively. A feasible approach is to treat the mass of the ground-state positive-parity hadron as a known input and subtract its contribution directly from the spectral density. Moreover, since the interpolating current in Eq.~(\ref{eq:12}) provides a good description of the $\Lambda_Q[\Xi_Q]$ baryon, we can construct a more complete interpolating current for the negative-parity state by changing its Dirac structure. In the framework of QCDSR, there are two independent Dirac structures that can couple to the $J^P=1/2^-$ states. The first one is obtained by directly multiplying the positive-parity current by the Dirac matrix $\gamma_5$, yielding $J_1(x) = \varepsilon_{ijk}u^{iT}(x) \mathcal{C} \gamma_5 d[s]^j(x)\gamma_5 Q^k(x)$. The second one is obtained by altering the Dirac structure of the light diquark, i.e., replacing $\mathcal{C}\gamma_5$ with $\mathcal{C}$, which gives $J_2(x) = \varepsilon_{ijk}u^{iT}(x) \mathcal{C} d[s]^j(x)Q^k(x)$. Since both currents share the same quantum numbers, a more general interpolating current can be constructed as a linear combination of them:
\begin{eqnarray}\label{eq:22}
	\notag
	J_{\Lambda^*_Q[\Xi^*_Q]}(x)&=&\cos\theta J_1(x)+\sin\theta J_2(x) \\
	\notag
	&=&\cos\theta\varepsilon_{ijk}u^{iT}(x)\mathcal{C}\gamma_5d[s]^j(x)\gamma_5Q^{k}(x)\\
	&+&\sin\theta\varepsilon_{ijk}u^{iT}(x)\mathcal{C}d[s]^j(x)Q^{k}(x),
\end{eqnarray}
where $\theta\in[0,2\pi]$ denotes the mixing angle. By tuning $\theta$, we can optimize the spectral density to suppress the continuum contamination and extract the optimal current for the negative-parity baryons.

Similar to the procedure for handling the positive-parity baryon, we first write down the the two-point correlation function for negative-parity baryon at the hadron side,
\begin{eqnarray}\label{eq:23}
	\notag
	\Pi_-(p)&&=\lambda_{\Lambda^*_Q[\Xi^*_Q]}^2\frac{\slashed{p}+m_{\Lambda^*_Q[\Xi^*_Q]}}{m_{\Lambda^*_Q[\Xi^*_Q]}^2-p^2}+\lambda_{\Lambda_Q[\Xi_Q]}^2\frac{\slashed{p}-m_{\Lambda_Q[\Xi_Q]}}{m_{\Lambda_Q[\Xi_Q]}^2-p^2}+h.c.\\
	\notag
	&&=\left(\frac{\lambda_{\Lambda^*_Q[\Xi^*_Q]}^2}{m_{\Lambda^*_Q[\Xi^*_Q]}^2-p^2}+\frac{\lambda_{\Lambda_Q[\Xi_Q]}^2}{m_{\Lambda_Q[\Xi_Q]}^2-p^2} \right)\slashed{p}\\
	&&+\left(\frac{\lambda_{\Lambda^*_Q[\Xi^*_Q]}^2m_{\Lambda^*_Q[\Xi^*_Q]}}{m_{\Lambda^*_Q[\Xi^*_Q]}^2-p^2}-\frac{\lambda_{\Lambda_Q[\Xi_Q]}^2m_{\Lambda_Q[\Xi_Q]}}{m_{\Lambda_Q[\Xi_Q]}^2-p^2} \right)+h.c.
\end{eqnarray}
Substituting the explicit form of the interpolating current for negative-parity baryons into Eq.~(\ref{eq:2}) and performing the OPE using Wick's theorem, we obtain the correlation function for negative-parity baryon at the QCD side:
\begin{eqnarray}\label{eq:24}
	\notag
	\Pi^{\rm{QCD}}_-(p)&=&i\varepsilon_{ijk}\varepsilon_{i'j'k'}\int d^4xe^{ip\cdot x}\\
	\notag
	&&\times\Big\{\cos^2\theta\mathrm{Tr}\{\gamma_5\mathcal{C}U^{ik'T}(x)\mathcal{C}\gamma_5D[S]^{jj'}(x)\}\gamma_5Q^{ki'}(x)\gamma_5\\
	\notag
	&&+\cos\theta\sin\theta\mathrm{Tr}\{\mathcal{C}U^{ik'T}(x)\mathcal{C}\gamma_5D[S]^{jj'}(x)\}\gamma_5Q^{ki'}(x)\\
	\notag
	&&+\cos\theta\sin\theta\mathrm{Tr}\{\mathcal{C}U^{ik'T}(x)\mathcal{C}D[S]^{jj'}(x)\gamma_5\}Q^{ki'}(x)\gamma_5\\
	&&+\sin^2\theta\mathrm{Tr}\{\mathcal{C}U^{ik'T}(x)\mathcal{C}D[S]^{jj'}(x)\}Q^{ki'}(x)\Big\},
\end{eqnarray}
After substituting the explicit form of the full propagator in Eq.~(\ref{eq:15}), we find that the cross terms proportional to $\cos\theta\sin\theta$ vanish. Consequently, the correlation function depends on $\theta$ only through $\cos^2\theta$ and $\sin^2\theta$, and is therefore periodic in $\theta$ with period $\pi$. This allows the mixing angle to be restricted to the range $[0,\pi]$. Furthermore, since $\sin^2(\pi-\theta)=\sin^2\theta$ and $\cos^2(\pi-\theta)=\cos^2\theta$, we have $\Pi^{\mathrm{QCD}}_-(\pi-\theta)=\Pi^{\mathrm{QCD}}_-(\theta)$. Therefore, the mixing angle can be further restricted to the range $[0,0.5\pi]$.

By matching the correlation functions of negative-parity baryon at both hadron and QCD side with quark-hadron duality condition, taking the mass of the positive-parity baryon obtained above as an input parameter, and performing the Borel transformation for the variable $P^2=-p^2$, we can obtain the following QCD sum rule for negative-parity baryon:
\begin{eqnarray}\label{eq:25}
	\notag
	&&\lambda_{\Lambda^*_Q[\Xi^*_Q]}^2(m_{\Lambda^*_Q[\Xi^*_Q]}+m_{\Lambda_Q[\Xi_Q]})\exp\left(-\frac{m_{\Lambda^*_Q[\Xi^*_Q]}^2}{T^2}\right)\\
	\notag
	&&=\int_{s_{\min}}^{s_0}ds\left[m_{\Lambda_Q[\Xi_Q]}\rho^{\rm{1QCD}}_-(s)+\rho^{2\rm{QCD}}_-(s)\right]\exp\left(-\frac{s}{T^2}\right).\\
\end{eqnarray}
Finally, the sum rule equations for the mass and pole residue of negative-parity baryon can be obtained as:
\begin{eqnarray}\label{eq:26}
	\notag
	m_{\Lambda^*_Q[\Xi^*_Q]}^2 &&= \frac{\int_{s_{\min}}^{s_0} dse^{-s/T^2} s\left[m_{\Lambda_Q[\Xi_Q]}\rho^{1\mathrm{QCD}}_-(s) + \rho^{2\mathrm{QCD}}_-(s)\right]}{\int_{s_{\min}}^{s_0} dse^{-s/T^2}\left[m_{\Lambda_Q[\Xi_Q]}\rho^{1\mathrm{QCD}}_-(s) + \rho^{2\mathrm{QCD}}_-(s)\right]},\\
	\notag
	\lambda _{\Lambda^*_Q[\Xi^*_Q]}^2 &&= \frac{\int_{s_{\min}}^{s_0} dse^{-s/T^2} \left[m_{\Lambda_Q[\Xi_Q]}\rho^{1\mathrm{QCD}}_-(s) + \rho^{2\mathrm{QCD}}_-(s)\right]}{(m_{\Lambda_Q[\Xi_Q]}+m_{\Lambda^*_Q[\Xi^*_Q]})e^{-m_{\Lambda^*_Q[\Xi^*_Q]}^2/T^2}}.\\
\end{eqnarray}
\section{Three-point QCD sum rules for the form factors of $\Lambda_b[\Xi_b]\to\Lambda^*_c[\Xi^*_c]$ transition}\label{sec3}
In this section, we use the interpolating currents of positive- and negative-parity baryons from the previous section, along with the corresponding masses and pole residues, to analyze the electroweak form factors of the transition processes $\Lambda_b\to\Lambda^*_c$ and $\Xi_b\to\Xi^*_c$. 

To analyze the transition form factors, we firstly write down the following three-point correlation function:
\begin{eqnarray}\label{eq:27}
	\notag
	\Pi_\mu(p,p')&=&i^2\int\int d^4xd^4ye^{ip'\cdot x}e^{iq\cdot y}\\
	&&\times\langle0|\mathcal{T}[J_{\Lambda^*_c[\Xi^*_c]}(x)\tilde{J}_\mu(y)\bar{J}_{\Lambda_b[\Xi_b]}(0)]|0\rangle,
\end{eqnarray}
where $p=p'+q$, $\tilde{J}_{\mu}$ is electroweak transition current, which has the following forms,
\begin{eqnarray}\label{eq:28}
	\tilde{J}_\mu(y)&&=\bar{c}(y)\gamma_\mu(1-\gamma_5) b(y),
\end{eqnarray}
At the hadron side, the complete sets of hadronic state coupled to corresponding interpolating currents is inserted into the correlation function. After performing the integrals in coordinate space, isolating the contributions of the ground states from excited states and using the double dispersion relation, the three-point correlation function can be represented as the following form~\cite{Reinders:1984sr,Colangelo:2000dp}:
\begin{widetext}
\begin{eqnarray}\label{eq:29}
		\notag
		\Pi _\mu^H(p,p') &&= \frac{\left\langle 0 \right|J_{\Lambda^*_c[\Xi^*_c]}(0) \left| \Lambda^*_c[\Xi^*_c](p') \right\rangle \left\langle \Lambda^*_c[\Xi^*_c](p') \right|\tilde{J}_\mu(0)\left| \Lambda _b[\Xi _b](p)\right\rangle \left\langle \Lambda _b[\Xi _b](p) \right|\bar J_{\Lambda _b[\Xi _b]}(0) \left| 0 \right\rangle}{(m_{\Lambda^*_c[\Xi^*_c]}^2-p'^2)(m_{\Lambda _b[\Xi _b]}^2-p^2)}\\
		\notag
		&&+ \frac{\left\langle 0 \right|J_{\Lambda^*_c[\Xi^*_c]}(0)\left| \Lambda _c[\Xi _c](p') \right\rangle \left\langle \Lambda _c[\Xi _c](p') \right|\tilde{J}_\mu(0)\left| \Lambda _b[\Xi _b](p) \right\rangle \left\langle \Lambda _b[\Xi _b](p) \right|\bar J_{\Lambda _b[\Xi _b]}(0)\left| 0 \right\rangle }{(m_{\Lambda _c[\Xi _c]}^2-p'^2)(m_{\Lambda _b[\Xi _b]}^2-p^2)}\\
		\notag
		&&+ \frac{\left\langle 0 \right|J_{\Lambda^*_c[\Xi^*_c]}(0)\left| \Lambda^*_c[\Xi^*_c](p') \right\rangle \left\langle \Lambda^*_c[\Xi^*_c](p') \right|\tilde{J}_\mu(0)\left| \Lambda _b^*[\Xi _b^*](p) \right\rangle \left\langle \Lambda _b^*[\Xi _b^*](p) \right|\bar J_{\Lambda _b[\Xi _b]}(0)\left| 0 \right\rangle }{(m_{\Lambda^*_c[\Xi^*_c]}^2-p'^2)(m_{\Lambda _b^*[\Xi _b^*]}^2-p^2)}\\
		&&+ \frac{\left\langle 0 \right|J_{\Lambda^*_c[\Xi^*_c]}(0)\left| \Lambda _c[\Xi _c](p') \right\rangle \left\langle \Lambda _c[\Xi _c](p') \right|\tilde{J}_\mu(0)\left|\Lambda _b^*[\Xi _b^*](p)\right\rangle \left\langle \Lambda _b^*[\Xi _b^*](p) \right|\bar J_{\Lambda _b[\Xi _b]}(0)\left| 0 \right\rangle}{(m_{\Lambda _c[\Xi _c]}^2-p'^2)(m_{\Lambda _b^*[\Xi _b^*]}^2-p^2)} + h.c.,
\end{eqnarray}
the hadron vacuum matrix elements in above equation can be defined as the pole residues in Eqs.~(\ref{eq:4}) and (\ref{eq:5}), the hadron transition matrix elements can be parameterized as the following form factors,
\begin{eqnarray}\label{eq:30}
	\notag
	\left\langle \Lambda^*_c[\Xi^*_c](p') \right|\tilde{J}_\mu(0)\left| \Lambda _b[\Xi _b](p) \right\rangle  &&= \bar u_{\Lambda^*_c[\Xi^*_c]}(p',s')\left[f^{+-}_1(q^2)\gamma_\mu+\frac{i\sigma_{\mu\nu}q^\nu}{m_{\Lambda_b[\Xi_b]}}f^{+-}_2(q^2)+\frac{q_\mu}{m_{\Lambda_b[\Xi_b]}}f^{+-}_3(q^2) \right]\gamma_5u_{\Lambda_b[\Xi_b]}(p,s),\\
	\notag
	&&-\bar u_{\Lambda^*_c[\Xi^*_c]}(p',s')\left[g^{+-}_1(q^2)\gamma_\mu+\frac{i\sigma_{\mu\nu}q^\nu}{m_{\Lambda_b[\Xi_b]}}g^{+-}_2(q^2)+\frac{q_\mu}{m_{\Lambda_b[\Xi_b]}}g^{+-}_3(q^2) \right]u_{\Lambda_b[\Xi_b]}(p,s),\\
	\notag
	\left\langle \Lambda_c[\Xi_c](p') \right|\tilde{J}_\mu(0)\left| \Lambda _b[\Xi _b](p) \right\rangle  &&= \bar u_{\Lambda_c[\Xi_c]}(p',s')\left[f^{++}_1(q^2)\gamma_\mu+\frac{i\sigma_{\mu\nu}q^\nu}{m_{\Lambda_b[\Xi_b]}}f^{++}_2(q^2)+\frac{q_\mu}{m_{\Lambda_b[\Xi_b]}}f^{++}_3(q^2) \right]u_{\Lambda_b[\Xi_b]}(p,s),\\
	\notag
	&&-\bar u_{\Lambda_c[\Xi_c]}(p',s')\left[g^{++}_1(q^2)\gamma_\mu+\frac{i\sigma_{\mu\nu}q^\nu}{m_{\Lambda_b[\Xi_b]}}g^{++}_2(q^2)+\frac{q_\mu}{m_{\Lambda_b[\Xi_b]}}g^{++}_3(q^2) \right]\gamma_5u_{\Lambda_b[\Xi_b]}(p,s),\\
	\notag
	\left\langle \Lambda^*_c[\Xi^*_c](p') \right|\tilde{J}_\mu(0)\left| \Lambda^*_b[\Xi^*_b](p) \right\rangle  &&= \bar u_{\Lambda^*_c[\Xi^*_c]}(p',s')\left[f^{--}_1(q^2)\gamma_\mu+\frac{i\sigma_{\mu\nu}q^\nu}{m_{\Lambda^*_b[\Xi^*_b]}}f^{--}_2(q^2)+\frac{q_\mu}{m_{\Lambda^*_b[\Xi^*_b]}}f^{--}_3(q^2) \right]u_{\Lambda^*_b[\Xi^*_b]}(p,s),\\
	\notag
	&&-\bar u_{\Lambda^*_c[\Xi^*_c]}(p',s')\left[g^{--}_1(q^2)\gamma_\mu+\frac{i\sigma_{\mu\nu}q^\nu}{m_{\Lambda^*_b[\Xi^*_b]}}g^{--}_2(q^2)+\frac{q_\mu}{m_{\Lambda^*_b[\Xi^*_b]}}g^{--}_3(q^2) \right]\gamma_5u_{\Lambda^*_b[\Xi^*_b]}(p,s),\\
	\notag
	\left\langle \Lambda_c[\Xi_c](p') \right|\tilde{J}_\mu(0)\left| \Lambda^*_b[\Xi^*_b](p) \right\rangle  &&= \bar u_{\Lambda_c[\Xi_c]}(p',s')\left[f^{-+}_1(q^2)\gamma_\mu+\frac{i\sigma_{\mu\nu}q^\nu}{m_{\Lambda^*_b[\Xi^*_b]}}f^{-+}_2(q^2)+\frac{q_\mu}{m_{\Lambda^*_b[\Xi^*_b]}}f^{-+}_3(q^2) \right]\gamma_5u_{\Lambda^*_b[\Xi^*_b]}(p,s),\\
	&&-\bar u_{\Lambda_c[\Xi_c]}(p',s')\left[g^{-+}_1(q^2)\gamma_\mu+\frac{i\sigma_{\mu\nu}q^\nu}{m_{\Lambda^*_b[\Xi^*_b]}}g^{-+}_2(q^2)+\frac{q_\mu}{m_{\Lambda^*_b[\Xi^*_b]}}g^{-+}_3(q^2) \right]u_{\Lambda^*_b[\Xi^*_b]}(p,s),
\end{eqnarray}
where $f_i$ and $g_i$ are vector and axial vector form factors, the superscript of form factors denotes the parity of the initial- and final-state hadrons in the transition matrix elements. For example, $+-$ and $-+$ denote the $\Lambda_b[\Xi_b]\to\Lambda^*_c[\Xi^*_c]$ and $\Lambda^*_b[\Xi^*_b]\to\Lambda_c[\Xi_c]$ transitions, respectively. Substituting the matrix
elements in Eq.~(\ref{eq:29}) with pole residues and form factors, the correlation functions at hadron side can be written as:
\begin{eqnarray}\label{eq:31}
	\notag
	\Pi _\mu ^{H}(p,p') &&= \frac{\lambda_{\Lambda^*_c[\Xi^*_c]}\lambda_{\Lambda_b[\Xi _b]}(\slashed p' + m_{\Lambda^*_c[\Xi^*_c]})\left\{ \begin{array}{l}
			\left[f^{+-}_1(q^2)\gamma_\mu+\frac{i\sigma_{\mu\nu}q^\nu}{m_{\Lambda_b[\Xi_b]}}f^{+-}_2(q^2)+\frac{q_\mu}{m_{\Lambda_b[\Xi_b]}}f^{+-}_3(q^2) \right]\gamma _5\\
			- \left[g^{+-}_1(q^2)\gamma_\mu+\frac{i\sigma_{\mu\nu}q^\nu}{m_{\Lambda_b[\Xi_b]}}g^{+-}_2(q^2)+\frac{q_\mu}{m_{\Lambda_b[\Xi_b]}}g^{+-}_3(q^2) \right]
		\end{array} \right\}(\slashed p +m_{\Lambda _b[\Xi _b]})}{(m_{\Lambda^*_c[\Xi^*_c]}^2-p'^2)(m_{\Lambda _b[\Xi _b]}^2-p^2)}\\
	\notag
	&&+ \frac{\lambda_{\Lambda_c[\Xi _c]}\lambda _{\Lambda _b[\Xi _b]}\gamma _5(\slashed p' + m_{\Lambda _c[\Xi _c]})\left\{\begin{array}{l}
		\left[f^{++}_1(q^2)\gamma_\mu+\frac{i\sigma_{\mu\nu}q^\nu}{m_{\Lambda_b[\Xi_b]}}f^{++}_2(q^2)+\frac{q_\mu}{m_{\Lambda_b[\Xi_b]}}f^{++}_3(q^2) \right]\\
			- \left[g^{++}_1(q^2)\gamma_\mu+\frac{i\sigma_{\mu\nu}q^\nu}{m_{\Lambda_b[\Xi_b]}}g^{++}_2(q^2)+\frac{q_\mu}{m_{\Lambda_b[\Xi_b]}}g^{++}_3(q^2) \right]\gamma_5
		\end{array} \right\}(\slashed p + m_{\Lambda _b[\Xi _b]})}{(m_{\Lambda _c[\Xi _c]}^2-p'^2)(m_{\Lambda _b[\Xi _b]}^2-p^2)}\\
	\notag
	&&- \frac{\lambda_{\Lambda^*_c[\Xi^*_b]}\lambda _{\Lambda _b^*[\Xi _b^*]}(\slashed p' + m_{\Lambda^*_c[\Xi^*_b]})\left\{ \begin{array}{l}
			\left[f^{--}_1(q^2)\gamma_\mu+\frac{i\sigma_{\mu\nu}q^\nu}{m_{\Lambda^*_b[\Xi^*_b]}}f^{--}_2(q^2)+\frac{q_\mu}{m_{\Lambda^*_b[\Xi^*_b]}}f^{--}_3(q^2) \right]\\
			- \left[g^{--}_1(q^2)\gamma_\mu+\frac{i\sigma_{\mu\nu}q^\nu}{m_{\Lambda^*_b[\Xi^*_b]}}g^{--}_2(q^2)+\frac{q_\mu}{m_{\Lambda^*_b[\Xi^*_b]}}g^{--}_3(q^2) \right]\gamma _5
		\end{array} \right\}(\slashed p + m_{\Lambda _b^*[\Xi _b^*]})\gamma _5}{(m_{\Lambda^* _c[\Xi^*_c]}^2-p'^2)(m_{\Lambda _b^*[\Xi _b^*]}^2-p^2)}\\
	\notag
	&&- \frac{\lambda _{\Lambda_c[\Xi _c]}\lambda _{\Lambda _b^*[\Xi _b^*]}\gamma _5(\slashed p' +m_{\Lambda _c[\Xi _c]})\left\{ \begin{array}{l}
			\left[f^{-+}_1(q^2)\gamma_\mu+\frac{i\sigma_{\mu\nu}q^\nu}{m_{\Lambda^*_b[\Xi^*_b]}}f^{-+}_2(q^2)+\frac{q_\mu}{m_{\Lambda^*_b[\Xi^*_b]}}f^{-+}_3(q^2) \right]\gamma _5\\
			- \left[g^{-+}_1(q^2)\gamma_\mu+\frac{i\sigma_{\mu\nu}q^\nu}{m_{\Lambda^*_b[\Xi^*_b]}}g^{-+}_2(q^2)+\frac{q_\mu}{m_{\Lambda^*_b[\Xi^*_b]}}g^{-+}_3(q^2) \right]
		\end{array} \right\}(\slashed p + m_{\Lambda _b^*[\Xi _b^*]})\gamma _5}{(m_{\Lambda _c[\Xi _c]}^2-p'^2)(m_{\Lambda _b^*[\Xi _b^*]}^2-p^2)} + ...\\
\end{eqnarray}
\end{widetext}
The above correlation function can be expanded into the following twenty-four independent dirac structures:
\begin{eqnarray}
	\notag
	\Pi^H_\mu(p,p')&&=\sum_{i=1}^{24}\Pi^H_i\mathrm{ST}_i,\\
	\notag
	\mathrm{ST}_i&&=\big\{\gamma_\mu\gamma_5, \gamma_\mu\slashed p'\gamma_5, \gamma_\mu\slashed q\gamma_5,  \gamma_\mu\slashed p'\slashed q\gamma_5, \slashed p'\gamma_5p'_\mu, \slashed p'\gamma_5q_\mu, \\
	\notag
	&&\slashed q\gamma_5p'_\mu, \slashed q\gamma_5q_\mu, \slashed p'\slashed q\gamma_5p'_\mu, \slashed p'\slashed q\gamma_5q_\mu, \gamma_5p'_\mu, \gamma_5q_\mu, \gamma_\mu,\\
	\notag
	&&\gamma_\mu\slashed p', \gamma_\mu\slashed q, \gamma_\mu\slashed p'\slashed q, \slashed p'p'_\mu, \slashed p'q_\mu, \slashed qp'_\mu, \slashed qq_\mu, \slashed p'\slashed qp'_\mu,\\
	&& \slashed p'\slashed qq_\mu,p'_\mu,q_\mu\big\}.
\end{eqnarray}
The twenty-four form factors defined by Eq.~(\ref{eq:30}) are included in expansion coefficients $\Pi^H_i(i=1,...,24)$, which are commonly called as the scalar invariant amplitudes. In the present work, we only focus on the form factors of $\Lambda_b[\Xi_b]\to\Lambda^*_c[\Xi^*_c]$ transition.

At the QCD side, we bring the specific form of interpolating and transition currents into the correlation function in Eq.~(\ref{eq:27}), and doing the OPE by Wick's theorem. Then, the correlation function at QCD side can be expressed as the following form:
\begin{eqnarray}\label{eq:33}
	\notag
	\Pi _\mu ^{\mathrm{QCD}}(p,p') &&= \varepsilon _{ijk}\varepsilon _{i'j'k'}\int d^4x d^4ye^{ip'x}e^{iqy} \\
	\notag
	&&\times\Big\{\cos\theta\mathrm{Tr}\big\{D[S]^{k'j}(x) \gamma _5\mathcal{C}U^{ki'T}(x)\mathcal{C}\gamma _5 \big\}\\
	\notag
	&&\times\gamma_5C^{j'm}(x - y)\gamma _\mu (1 - \gamma _5)B^{mi}(y)\\
	\notag
	&&+\sin\theta\mathrm{Tr}\big\{D[S]^{k'j}(x)\gamma_5\mathcal{C}U^{ki'T}(x)\mathcal{C}\big\}\\
	&&\times C^{j'm}(x - y)\gamma _\mu (1 - \gamma _5)B^{mi}(y)\Big\}.
\end{eqnarray} 
Similarly to the hadron side, the correlation functions at the QCD side can also be expanded into the following scalar invariant amplitudes,
\begin{eqnarray}
	\Pi^{\mathrm{QCD}}_\mu(p,p')&&=\sum_{i=1}^{24}\Pi^{\mathrm{QCD}}_i\mathrm{ST}_i,
\end{eqnarray} 
the scalar invariant amplitudes at QCD side can be written as the following double dispersion integration,
\begin{eqnarray}
	\Pi _i ^{\mathrm{QCD}}(p,p') = \int_{{u_{\min }}}^\infty  du \int_{{s_{\min }}}^\infty  ds \frac{\rho _i ^{\mathrm{QCD}}(s,u,q^2)}{(s - p^2)(u - p'^2)}.
\end{eqnarray}
Here $\rho^{\mathrm{QCD}}_i(s,u,q^2)$ is the QCD spectral density with $s=p^2$ and $u=p'^2$, and can be written as:
\begin{eqnarray}
	\rho_i ^{\mathrm{QCD}}(s,u,q^2) =\frac{1}{(2\pi i)^2}\mathrm{Dis}_s\mathrm{Dis}_u\Pi_i ^{\mathrm{QCD}}(s,u,q^2),
\end{eqnarray}
The QCD spectral density can be expressed as the summation of perturbative part and different vacuum condensate terms:
\begin{eqnarray}
	\notag
	\rho _i ^{\mathrm{QCD}}(s,u,q^2) &&=\rho_i^{pert}(s,u,q^2)+\rho_i^{\langle \bar{q}q\rangle}(s,u,q^2)\\
	\notag
	&&+\rho_i^{\langle \bar{s}s\rangle}(s,u,q^2)+\rho_i^{\langle g^2_sGG\rangle}(s,u,q^2)\\
	\notag
	&&+\rho_i^{\langle \bar{q}g_s\sigma Gq\rangle}(s,u,q^2)+\rho_i^{\langle \bar{s}g_s\sigma Gs\rangle}(s,u,q^2)\\
	\notag
	&&+\rho_i^{\langle \bar{q}q\rangle^2}(s,u,q^2)+\rho_i^{\langle \bar{q}q\rangle \langle \bar{s}s\rangle}(s,u,q^2)\\
	\notag
	&&+\rho_i^{\langle \bar{q}q\rangle\langle g_s^2GG\rangle}(s,u,q^2)+\rho_i^{\langle \bar{s}s\rangle\langle g_s^2GG\rangle}(s,u,q^2)\\
	\notag
	&&+\rho_i^{\langle \bar{q}q\rangle\langle \bar{q}g_s\sigma Gq\rangle}(s,u,q^2)+\rho_i^{\langle \bar{q}q\rangle\langle \bar{s}g_s\sigma Gs\rangle}(s,u,q^2)\\
	&&+\rho_i^{\langle \bar{s}s\rangle\langle \bar{q}g_s\sigma Gq\rangle}(s,u,q^2),
\end{eqnarray}
In the present work, we truncate the OPE up to a maximum of 8 dimensions, with the corresponding Feynman diagrams similar to those in Ref.~\cite{Lu:2026qkk}. The calculation details can be found in Refs.~\cite{Lu:2025gol,Yu:2026tbk}.

Taking the variable change $p^2\to-P^2$, $p'^2\to-P'^2$ and $q^2\to-Q^2$, and doing the double Borel transforms for variables $P^2$ and $P'^2$ to the correlation functions at both hadron and QCD sides~\cite{Reinders:1984sr}. The variables $P^2$ and $P'^2$ will be replaced by Borel parameters $T_1^2$ and $T_2^2$, respectively. In this work, the relations $k=T_2^2/T_1^2=m_{\Lambda^*_c[\Xi^*_c]}^2/m_{\Lambda_b[\Xi_b]}^2$ and $T^2=T_1^2$ are adopted to reduce the Borel parameters. Then, according to the quark-hadron duality condition, a series of linear equations about twenty-four scalar invariant amplitudes can be established, and all form factors in Eq.~(\ref{eq:31}) can be uniquely determined by solving these twenty-four linear equations. In this article, we only focus on the form factors for the $\Lambda_b[\Xi_b]\to\Lambda^*_c[\Xi^*_c]$ transition. The QCD sum rules for corresponding form factors can be represented as the following forms:
\begin{widetext}
\begin{eqnarray}\label{eq:38}
        \notag
		f_1^{+-}(Q^2) &&= \frac{e^{m^2_{\Lambda _b[\Xi _b]}/T^2 + m^2_{\Lambda _c[\Xi _c]}/kT^2}}{2\lambda _{\Lambda^*_c[\Xi^*_c]}{\lambda _{\Lambda _b[\Xi _b]}}(m_{\Lambda _c[\Xi _c]} + m_{\Lambda _c^*[\Xi _c^*]})(m_{\Lambda _b[\Xi _b]}+ m_{\Lambda _b^*[\Xi _b^*]})}\int_{{u_{\min }}}^{u_0} du \int_{{s_{\min }}}^{s_0} ds e^{ - s/T^2 - u/kT^2}\\
		\notag
		&&\times \left\{ \begin{array}{l} 
					2\rho _1^{\mathrm{QCD}}(s,u,Q^2) - 2(m_{\Lambda _b[\Xi _b]} + m_{\Lambda _c[\Xi _c]}-m_{\Lambda _c^*[\Xi _c^*]})\rho _2^{\mathrm{QCD}}(s,u,Q^2) + 2(m_{\Lambda _b[\Xi _b]} - m_{\Lambda _b^*[\Xi _b^*]})\rho _3^{\mathrm{QCD}}(s,u,Q^2)\\
					+ 2[m_{\Lambda _b[\Xi _b]}(m_{\Lambda _b^*[\Xi _b^*]} - m_{\Lambda _c[\Xi _c]} + m_{\Lambda _c^*[\Xi _c^*]}) + m_{\Lambda _b^*[\Xi _b^*]}(m_{\Lambda _c[\Xi _c]} - m_{\Lambda _c^*[\Xi _c^*]}) - m_{\Lambda _c^*[\Xi _c^*]}^2]\rho _4^{\mathrm{QCD}}(s,u,Q^2)\\
					+ m_{\Lambda _c[\Xi _c]}(m_{\Lambda _c^*[\Xi _c^*]} - m_{\Lambda _b[\Xi _b]})\rho_5^{\mathrm{QCD}}(s,u,Q^2) + (m_{\Lambda _b[\Xi _b]} - m_{\Lambda _c^*[\Xi _c^*]})(m_{\Lambda _b^*[\Xi _b^*]} + m_{\Lambda _c^*[\Xi _c^*]})\rho _7^{\mathrm{QCD}}(s,u,Q^2)\\
					+ m_{\Lambda _c[\Xi _c]}(m_{\Lambda _b[\Xi _b]} - m_{\Lambda _c^*[\Xi _c^*]})(m_{\Lambda _b^*[\Xi _b^*]} + m_{\Lambda _c^*[\Xi _c^*]})\rho _9^{\mathrm{QCD}}(s,u,Q^2)\\
					+ (m_{\Lambda _c^*[\Xi _c^*]} - m_{\Lambda _b[\Xi _b]})\rho _{11}^{\mathrm{QCD}}(s,u,Q^2)
				 \end{array} \right\},
\end{eqnarray}
\begin{eqnarray}
		\notag
		f_2^{+-}(Q^2) &&= \frac{m_{\Lambda _b[\Xi _b]}e^{m^2_{\Lambda _b[\Xi _b]}/T^2 + m^2_{\Lambda _c[\Xi _c]}/kT^2}}{2\lambda _{\Lambda _c[\Xi _c]}{\lambda _{\Lambda _b[\Xi _b]}}(m_{\Lambda _c[\Xi _c]} + m_{\Lambda _c^*[\Xi _c^*]})(m_{\Lambda _b[\Xi _b]} + m_{\Lambda _b^*[\Xi _b^*]})}\int_{{u_{\min }}}^{u_0} du \int_{{s_{\min }}}^{s_0} ds e^{-s/T^2 - u/kT^2}\\
		\notag
		&&\times \left\{ \begin{array}{l}
				2\rho _2^{\mathrm{QCD}}(s,u,Q^2) - 2\rho _3^{\mathrm{QCD}}(s,u,Q^2) - 2({m_{\Lambda _b^*[\Xi _b^*]}} - m_{\Lambda _c[\Xi _c]} + m_{\Lambda _c^*[\Xi _c^*]})\rho _4^{\mathrm{QCD}}(s,u,Q^2) + m_{\Lambda _c[\Xi _c]}\rho _5^{\mathrm{QCD}}(s,u,Q^2)\\
				- (m_{\Lambda _b^*[\Xi _b^*]} + m_{\Lambda _c^*[\Xi _c^*]})\rho _7^{\mathrm{QCD}}(s,u,Q^2) - m_{\Lambda _c[\Xi _c]}(m_{\Lambda _b^*[\Xi _b^*]} + m_{\Lambda _c^*[\Xi _c^*]})\rho _9^{\mathrm{QCD}}(s,u,Q^2) + \rho _{11}^{\mathrm{QCD}}(s,u,Q^2)
		\end{array} \right\},
\end{eqnarray}
\begin{eqnarray}
		\notag
		f_3^{+-}(Q^2) &&= \frac{m_{\Lambda _b[\Xi _b]}e^{m^2_{\Lambda _b[\Xi _b]}/T^2 + m^2_{\Lambda _c[\Xi _c]}/kT^2}}{2\lambda _{\Lambda _c[\Xi _c]}\lambda _{\Lambda _b[\Xi _b]}(m_{\Lambda _c[\Xi _c]} + m_{\Lambda _c^*[\Xi _c^*]})(m_{\Lambda _b[\Xi _b]} + m_{\Lambda _b^*[\Xi _b^*]})}\int_{{u_{\min }}}^{u_0} du \int_{{s_{\min }}}^{s_0} ds e^{-s/T^2 - u/kT^2}\\
		\notag
		&&\times \left\{ \begin{array}{l}
				- 2\rho _2^{\mathrm{QCD}}(s,u,Q^2) + 2\rho _3^{\mathrm{QCD}}(s,u,Q^2) + 2(m_{\Lambda _b^*[\Xi _b^*]} - m_{\Lambda _c[\Xi _c]} + m_{\Lambda _c^*[\Xi _c^*]})\rho _4^{\mathrm{QCD}}(s,u,Q^2) - m_{\Lambda _c[\Xi _c]}\rho _5^{\mathrm{QCD}}(s,u,Q^2)\\
				+ 2m_{\Lambda _c[\Xi _c]}\rho _6^{\mathrm{QCD}}(s,u,Q^2) + (m_{\Lambda _b^*[\Xi _b^*]} + m_{\Lambda _c^*[\Xi _c^*]})\rho _7^{\mathrm{QCD}}(s,u,Q^2) - 2(m_{\Lambda _b^*[\Xi _b^*]} + m_{\Lambda _c^*[\Xi _c^*]})\rho _8^{\mathrm{QCD}}(s,u,Q^2)\\
				+ m_{\Lambda _c[\Xi _c]}(m_{\Lambda _b^*[\Xi _b^*]} + m_{\Lambda _c^*[\Xi _c^*]})\rho _9^{\mathrm{QCD}}(s,u,Q^2) - 2m_{\Lambda _c[\Xi _c]}(m_{\Lambda _b^*[\Xi _b^*]} + m_{\Lambda _c^*[\Xi _c^*]})\rho _{10}^{\mathrm{QCD}}(s,u,{Q^2})\\
				- \rho _{11}^{\mathrm{QCD}}(s,u,Q^2) + 2\rho _{12}^{\mathrm{QCD}}(s,u,Q^2)
		\end{array} \right\},
\end{eqnarray}
\begin{eqnarray}
	\notag
	g_1^{+-}(Q^2) &&= \frac{e^{m^2_{\Lambda _b[\Xi _b]}/T^2 + m^2_{\Lambda _c[\Xi _c]}/kT^2}}{2\lambda _{\Lambda^*_c[\Xi^*_c]}{\lambda _{\Lambda _b[\Xi _b]}}(m_{\Lambda _c[\Xi _c]} + m_{\Lambda _c^*[\Xi _c^*]})(m_{\Lambda _b[\Xi _b]}+ m_{\Lambda _b^*[\Xi _b^*]})}\int_{{u_{\min }}}^{u_0} du \int_{{s_{\min }}}^{s_0} ds e^{ - s/T^2 - u/kT^2}\\
	\notag
	&&\times \left\{ \begin{array}{l} 
		2\rho _{13}^{\mathrm{QCD}}(s,u,Q^2) + 2(m_{\Lambda _b[\Xi _b]} - m_{\Lambda _c[\Xi _c]}+m_{\Lambda _c^*[\Xi _c^*]})\rho _{14}^{\mathrm{QCD}}(s,u,Q^2) - 2(m_{\Lambda _b[\Xi _b]} - m_{\Lambda _b^*[\Xi _b^*]})\rho _{15}^{\mathrm{QCD}}(s,u,Q^2)\\
		+ 2[m_{\Lambda _b[\Xi _b]}(m_{\Lambda _b^*[\Xi _b^*]} + m_{\Lambda _c[\Xi _c]} - m_{\Lambda _c^*[\Xi _c^*]}) - m_{\Lambda _b^*[\Xi _b^*]}(m_{\Lambda _c[\Xi _c]} - m_{\Lambda _c^*[\Xi _c^*]}) - m_{\Lambda _c^*[\Xi _c^*]}^2]\rho _{16}^{\mathrm{QCD}}(s,u,Q^2)\\
		+ m_{\Lambda _c[\Xi _c]}(m_{\Lambda _c^*[\Xi _c^*]} + m_{\Lambda _b[\Xi _b]})\rho_{17}^{\mathrm{QCD}}(s,u,Q^2) + (m_{\Lambda _b[\Xi _b]} + m_{\Lambda _c^*[\Xi _c^*]})(m_{\Lambda _b^*[\Xi _b^*]} - m_{\Lambda _c^*[\Xi _c^*]})\rho _{19}^{\mathrm{QCD}}(s,u,Q^2)\\
		+ m_{\Lambda _c[\Xi _c]}(m_{\Lambda _b[\Xi _b]} + m_{\Lambda _c^*[\Xi _c^*]})(m_{\Lambda _b^*[\Xi _b^*]} - m_{\Lambda _c^*[\Xi _c^*]})\rho _{21}^{\mathrm{QCD}}(s,u,Q^2)\\
		+ (m_{\Lambda _c^*[\Xi _c^*]} + m_{\Lambda _b[\Xi _b]})\rho _{23}^{\mathrm{QCD}}(s,u,Q^2)
	\end{array} \right\},
\end{eqnarray}
\begin{eqnarray}
	\notag
	g_2^{+-}(Q^2) &&= \frac{m_{\Lambda _b[\Xi _b]}e^{m^2_{\Lambda _b[\Xi _b]}/T^2 + m^2_{\Lambda _c[\Xi _c]}/kT^2}}{2\lambda _{\Lambda _c[\Xi _c]}{\lambda _{\Lambda _b[\Xi _b]}}(m_{\Lambda _c[\Xi _c]} + m_{\Lambda _c^*[\Xi _c^*]})(m_{\Lambda _b[\Xi _b]} + m_{\Lambda _b^*[\Xi _b^*]})}\int_{{u_{\min }}}^{u_0} du \int_{{s_{\min }}}^{s_0} ds e^{-s/T^2 - u/kT^2}\\
	\notag
	&&\times \left\{ \begin{array}{l}
		2\rho _{14}^{\mathrm{QCD}}(s,u,Q^2) - 2\rho _{15}^{\mathrm{QCD}}(s,u,Q^2) + 2({m_{\Lambda _b^*[\Xi _b^*]}} + m_{\Lambda _c[\Xi _c]} - m_{\Lambda _c^*[\Xi _c^*]})\rho _{16}^{\mathrm{QCD}}(s,u,Q^2) - m_{\Lambda _c[\Xi _c]}\rho _{17}^{\mathrm{QCD}}(s,u,Q^2)\\
		+ (m_{\Lambda _b^*[\Xi _b^*]} - m_{\Lambda _c^*[\Xi _c^*]})\rho _{19}^{\mathrm{QCD}}(s,u,Q^2) + m_{\Lambda _c[\Xi _c]}(m_{\Lambda _b^*[\Xi _b^*]} - m_{\Lambda _c^*[\Xi _c^*]})\rho _{21}^{\mathrm{QCD}}(s,u,Q^2) + \rho _{23}^{\mathrm{QCD}}(s,u,Q^2)
	\end{array} \right\},
\end{eqnarray}
\begin{eqnarray}
	\notag
	g_3^{+-}(Q^2) &&= \frac{m_{\Lambda _b[\Xi _b]}e^{m^2_{\Lambda _b[\Xi _b]}/T^2 + m^2_{\Lambda _c[\Xi _c]}/kT^2}}{2\lambda _{\Lambda _c[\Xi _c]}\lambda _{\Lambda _b[\Xi _b]}(m_{\Lambda _c[\Xi _c]} + m_{\Lambda _c^*[\Xi _c^*]})(m_{\Lambda _b[\Xi _b]} + m_{\Lambda _b^*[\Xi _b^*]})}\int_{{u_{\min }}}^{u_0} du \int_{{s_{\min }}}^{s_0} ds e^{-s/T^2 - u/kT^2}\\
	\notag
	&&\times \left\{ \begin{array}{l}
		- 2\rho _{14}^{\mathrm{QCD}}(s,u,Q^2) + 2\rho _{15}^{\mathrm{QCD}}(s,u,Q^2) - 2(m_{\Lambda _b^*[\Xi _b^*]} + m_{\Lambda _c[\Xi _c]} - m_{\Lambda _c^*[\Xi _c^*]})\rho _{16}^{\mathrm{QCD}}(s,u,Q^2) - m_{\Lambda _c[\Xi _c]}\rho _{17}^{\mathrm{QCD}}(s,u,Q^2)\\
		+ 2m_{\Lambda _c[\Xi _c]}\rho _{18}^{\mathrm{QCD}}(s,u,Q^2) - (m_{\Lambda _b^*[\Xi _b^*]} - m_{\Lambda _c^*[\Xi _c^*]})\rho _{19}^{\mathrm{QCD}}(s,u,Q^2) + 2(m_{\Lambda _b^*[\Xi _b^*]} - m_{\Lambda _c^*[\Xi _c^*]})\rho _{20}^{\mathrm{QCD}}(s,u,Q^2)\\
		- m_{\Lambda _c[\Xi _c]}(m_{\Lambda _b^*[\Xi _b^*]} - m_{\Lambda _c^*[\Xi _c^*]})\rho _{21}^{\mathrm{QCD}}(s,u,Q^2) + 2m_{\Lambda _c[\Xi _c]}(m_{\Lambda _b^*[\Xi _b^*]} - m_{\Lambda _c^*[\Xi _c^*]})\rho _{22}^{\mathrm{QCD}}(s,u,{Q^2})\\
		- \rho _{23}^{\mathrm{QCD}}(s,u,Q^2) + 2\rho _{24}^{\mathrm{QCD}}(s,u,Q^2)
	\end{array} \right\},\\
\end{eqnarray}
\end{widetext}
Here, $s_0$ and $u_0$ are continuum threshold parameters for initial- and final-state baryons. For convenience, we ignore the superscript of the form factors below and use $f_i$ and $g_i$ to represent $f_i^{+-}$ and $g_i^{+-}$.
\section{Weak decay of $\Lambda_b[\Xi_b]\to\Lambda^*_c[\Xi^*_c]$}\label{sec4}
In this section, we analyze the semileptonic and nonleptonic decay processes related to $\Lambda_b[\Xi_b]\to\Lambda^*_c[\Xi^*_c]$ by using the electroweak form factors obtained in previous section.
\subsection{Semileptonic decay $\Lambda_b[\Xi_b]\to\Lambda^*_c[\Xi^*_c]\ell\bar{\nu}_\ell$}\label{sec4A}
The semileptonic decay $\Lambda_b[\Xi_b]\to\Lambda^*_c[\Xi^*_c]\ell\bar{\nu}_\ell$ ($\ell=e, \mu$ and $\tau$) can be described by the following effective Hamiltonian,
\begin{eqnarray}\label{eq:39}
	H_{eff} = \frac{G_F}{\sqrt 2}\sum\limits_{\ell = e,\mu ,\tau } {V_{cb}\bar c\gamma _\mu (1 - \gamma _5)b\bar \ell\gamma ^\mu(1 - \gamma _5)\nu_\ell} 
\end{eqnarray}
where $G_{F}=1.16637\times 10^{-5}$ GeV$^{-2}$ is the Fermi constant, $V_{cb}$ is CKM matrix element. The three-body differential decay distribution can be written as:
\begin{eqnarray}\label{eq:40}
	\frac{d^2\Gamma}{dq^2d\cos\theta_\ell}=\frac{G_F^2}{(2\pi)^3}|V_{cb}|^2\frac{\sqrt{Q_+Q_-}}{32m^3_{\Lambda_b[\Xi_b]}}\left(1-\frac{m_\ell^2}{q^2}\right)L^{\mu\nu}H_{\mu\nu},
\end{eqnarray}
where $q^2$ denote the four momentum squared of the intermediate virtual off-shell $W$ Boson, $Q_\pm=(m_{\Lambda_b[\Xi_b]}\pm m_{\Lambda^*_c[\Xi^*_c]})^2-q^2$, $\theta_\ell$ is the polar angle of the final state lepton, and $L^{\mu\nu}$ and $H_{\mu\nu}$ are lepton and hadron currents tensors, respectively, which have the following forms:
\begin{eqnarray}\label{eq:41}
	\notag
	L^{\mu\nu}&=&\langle \ell(l)\bar{\nu}_\ell(v)|\bar \ell\gamma ^\mu(1 - \gamma _5)\nu_\ell|0\rangle^*\\
	\notag
	&&\times\langle \ell(l)\bar{\nu}_\ell(v)|\bar \ell\gamma ^\nu(1 - \gamma _5)\nu_\ell|0\rangle, \\
	\notag
	H_{\mu\nu}&=&\langle \Lambda^*_c[\Xi^*_c](p')|\bar{c}\gamma _\mu(1 - \gamma _5)b|\Lambda_b[\Xi_b](p)\rangle^*\\
	&&\times\langle \Lambda^*_c[\Xi^*_c](p')|\bar{c}\gamma _\nu(1 - \gamma _5)b|\Lambda_b[\Xi_b](p)\rangle.
\end{eqnarray}
According the helicity amplitude and Wigner’s $d^J$–functions, the contraction of the lepton and hadron tensors $L^{\mu\nu}H_{\mu\nu}$ can be expressed as~\cite{Kadeer:2005aq},
\begin{eqnarray}\label{eq:42}
	\notag
	L^{\mu\nu}H_{\mu\nu}&=&\frac{1}{8}\sum_{\lambda_\ell,\lambda_2,\lambda_W,\lambda'_W,J,J'}(-1)^{J+J'}|h^{\ell}_{\lambda_\ell\lambda_{\bar{\nu}_\ell}=\frac{1}{2}}|^2\delta_{\lambda_2-\lambda_W,\lambda_2-\lambda_W'}\\
	&&\times d^J_{\lambda_W,\lambda_\ell=-\frac{1}{2}}(\theta_\ell)d^{J'}_{\lambda_W',\lambda_\ell=-\frac{1}{2}}(\theta_\ell)H_{\lambda_2\lambda_W}H^*_{\lambda_2\lambda'_W},
\end{eqnarray}
where $\lambda_2=\pm1/2$, $\lambda_\ell=\pm1/2$, $\lambda_{\bar{\nu}_\ell}=1/2$ and $\lambda_W=t,\pm1$ and $0$ are the helicity components of final state baryon, lepton, antineutrino and off-shell $W$ Boson, $J=1$ is for $\lambda_W=\pm1,0$ and $J=0$ for $\lambda_W=t$. The Wigner’s $d^J$–function matrix for $J=1$ is written as,
\begin{eqnarray}
	\notag
    d^1_{mm'}(\theta_\ell)=\left( {\begin{array}{*{20}{c}}
    		\frac{1}{2}(1+\cos\theta_\ell)&-\frac{1}{\sqrt{2}}\sin\theta_\ell&\frac{1}{2}(1-\cos\theta_\ell)\\
    		\frac{1}{\sqrt{2}}\sin\theta_\ell&\cos\theta_\ell&-\frac{1}{\sqrt{2}}\sin\theta_\ell\\
    		\frac{1}{2}(1-\cos\theta_\ell)&\frac{1}{\sqrt{2}}\sin\theta_\ell&\frac{1}{2}(1+\cos\theta_\ell)
    \end{array}} \right),
\end{eqnarray}
where the rows and columns are labeled in the order $(1,0,-1)$. The moduli squared of the lepton helicity amplitudes $|h^\ell_{\lambda_\ell\lambda_{\bar{\nu}_\ell}}|^2$ can be represented as:
\begin{eqnarray}\label{eq:43}
	\notag
	\mathrm{nonflip}(\lambda_W=\mp1):&&~|h^\ell_{\lambda_\ell=-\frac{1}{2}\lambda_{\bar{\nu}_\ell}=\frac{1}{2}}|^2=8(q^2-m_\ell^2),\\
	\mathrm{flip}(\lambda_W=t,0):&&~|h^\ell_{\lambda_\ell=\frac{1}{2}\lambda_{\bar{\nu}_\ell}=\frac{1}{2}}|^2=8\frac{m_\ell^2}{2q^2}(q^2-m_\ell^2).
\end{eqnarray}
The hadron helicity amplitudes can be expressed as follows,
\begin{eqnarray}\label{eq:44}
	\notag
	H_{\frac{1}{2},0}^V &&=\frac{\sqrt {Q_+}}{\sqrt{q^2}}\left[ (m_{\Lambda _b[\Xi _b]}- m_{\Lambda^*_c[\Xi^*_c]})f_1(q^2)+ \frac{q^2}{m_{\Lambda _b[\Xi _b]}}f_2(q^2) \right],\\
	\notag
	H_{\frac{1}{2},1}^V &&= \sqrt {2Q_+} \left[-f_1(q^2) - \frac{m_{\Lambda _b[\Xi _b]} - m_{\Lambda^*_c[\Xi^*_c]}}{m_{\Lambda _b[\Xi _b]}}f_2(q^2) \right],\\
	\notag
	H_{\frac{1}{2},t}^V &&= \frac{\sqrt {Q_ -}}{\sqrt{q^2}}\left[ (m_{\Lambda _b[\Xi _b]}+ m_{\Lambda^*_c[\Xi^*_c]})f_1(q^2) - \frac{q^2}{m_{\Lambda _b[\Xi _b]}}f_3(q^2) \right],
\end{eqnarray}
\begin{eqnarray}
	\notag
	H_{\frac{1}{2},0}^A &&=\frac{\sqrt {Q_-}}{\sqrt{q^2}}\left[ (m_{\Lambda _b[\Xi _b]}+ m_{\Lambda^*_c[\Xi^*_c]})g_1(q^2)- \frac{q^2}{m_{\Lambda _b[\Xi _b]}}g_2(q^2) \right],\\
    \notag
    H_{\frac{1}{2},1}^A &&= \sqrt {2Q_-} \left[-g_1(q^2) + \frac{m_{\Lambda _b[\Xi _b]} +m_{\Lambda^*_c[\Xi^*_c]}}{m_{\Lambda _b[\Xi _b]}}g_2(q^2) \right],\\
    \notag
    H_{\frac{1}{2},t}^A &&= \frac{\sqrt {Q_ +}}{\sqrt{q^2}}\left[ (m_{\Lambda _b[\Xi _b]}- m_{\Lambda^*_c[\Xi^*_c]})g_1(q^2) + \frac{q^2}{m_{\Lambda _b[\Xi _b]}}g_3(q^2) \right],\\
\end{eqnarray}
where the superscripts $V$ and $A$ denote vector and axial-vector helicity amplitudes, the negative and total helicity amplitudes can be obtained by
\begin{eqnarray}
	\notag
	H_{-\lambda _2, -\lambda _W}^V &=& H_{\lambda _2,\lambda _W}^V,\\
	\notag
	H_{-\lambda _2, -\lambda _W}^A &=&- H_{\lambda _2,\lambda _W}^A,\\
	H_{\lambda _2, \lambda _W}&=& H^V_{\lambda _2, \lambda _W}-H^A_{\lambda _2, \lambda _W}
\end{eqnarray}
Performing the sum in Eq.~(\ref{eq:42}) ($\lambda_2=\pm1/2$, $\lambda_\ell=\pm1/2$, $\lambda_W=t,\pm1,0$ and $J(J')=0,1$), the three-body differential decay distribution can obtained as:
\begin{eqnarray}\label{eq:46}
	\notag
	\frac{d^2\Gamma}{dq^2d\cos\theta_\ell}&&=\frac{G_F^2}{(2\pi)^3}|V_{cb}|^2\frac{\sqrt{Q_+Q_-}}{32m^3_{\Lambda_b[\Xi_b]}}\left(1-\frac{m_\ell^2}{q^2}\right)\frac{2}{3}(q^2-m_\ell^2)\\
	\notag
	&&\times\Big\{\frac{3}{8}(1-\cos\theta_\ell)^2|H_{\frac{1}{2},1}|^2+\frac{3}{8}(1+\cos\theta_\ell)^2|H_{-\frac{1}{2},-1}|^2\\
	\notag
	&&+\frac{3}{4}\sin\theta_\ell\big(|H_{\frac{1}{2},0}|^2+|H_{-\frac{1}{2},0}|^2\big)+\frac{m_\ell^2}{2q^2}\\
	\notag
	&&\times\Big[\frac{3}{2}\big(|H_{\frac{1}{2},t}|^2+|H_{-\frac{1}{2},t}|^2\big)\\
	\notag
	&&+\frac{3}{4}\sin^2\theta_\ell\big(|H_{\frac{1}{2},1}|^2+|H_{-\frac{1}{2},-1}|^2\big)\\
	\notag
	&&+\frac{3}{2}\cos^2\theta_\ell\big(|H_{\frac{1}{2},0}|^2+|H_{-\frac{1}{2},0}|^2\big)\\
	&&-3\cos\theta_\ell\big(H_{\frac{1}{2},t}H_{\frac{1}{2},0}+H_{-\frac{1}{2},t}H_{-\frac{1}{2},0}\big)\Big]\Big\}.
\end{eqnarray}
\subsection{Nonleptonic decay $\Lambda_b[\Xi_b]\to\Lambda^*_c[\Xi^*_c]P[V]$}\label{sec4B}
The nonleptonic decays $\Lambda_b[\Xi_b]\to\Lambda^*_c[\Xi^*_c]P$ and $\Lambda_b[\Xi_b]\to\Lambda^*_c[\Xi^*_c]V$, where $P$ denote the light pseudoscalar meson $\pi$ or $K$ and $V$ denote the light vector meson $\rho$ or $K^*$ can be uniformly described by the
following effective Hamiltonian:
\begin{eqnarray}\label{eq:47}
	H_{eff} = \frac{G_F}{\sqrt 2}\sum_{q=d,s} V_{cb}V^*_{uq}a_1\bar c\gamma _\mu (1 - \gamma _5)b\bar q\gamma ^\mu(1 - \gamma _5)u,
\end{eqnarray}
where $V_{cb}$ and $V_{uq}$ are CKM matrix elements, and $a_1$ is the effective Wilson coefficient. The decay width can be expressed by the following standard two-body decay formula, 
\begin{eqnarray}\label{eq:48}
	\notag
	\Gamma&=&\frac{\textbf{p}}{16\pi m_{\Lambda_b[\Xi_b]}^2}\sum|M|^2,\\
	\textbf{p}&=&\frac{\sqrt{\lambda\left(m_{\Lambda_b[\Xi_b]}^2,m_{\Lambda^*_c[\Xi^*_c]}^2,m_{P[V]}^2\right)}}{2 m_{\Lambda_b[\Xi_b]}},
\end{eqnarray}
here $\sum$ denotes the summation of all the spinor and polarization vectors and $\lambda(a,b,c)=a^2+b^2+c^2-2ab-2bc-2ac$ is the \text{K\"all\'en} function. The transition matrix for these decay processes can be expressed as the following form by using the factorization approach:
\begin{eqnarray}\label{eq:49}
	\notag
	M&&=\langle \Lambda^*_c[\Xi^*_c](p')P[V](q)|H_{eff}|\Lambda_b[\Xi_b](p)\rangle\\
	\notag
	&&=\frac{G_F}{\sqrt 2} V_{cb}V^*_{uq}a_1\langle \Lambda^*_c[\Xi^*_c](p')|\bar{c}\gamma _\mu(1 - \gamma _5)b|\Lambda_b[\Xi_b](p)\rangle\\
	&&\times\langle P[V](q)|\bar{q}\gamma _\mu(1 - \gamma _5)u|0\rangle,
\end{eqnarray}
the meson vacuum matrix elements can be defined as follows,
\begin{eqnarray}\label{eq:50}
 	\notag
 	\langle P(q)|\bar{q}\gamma^\mu(1 - \gamma _5)u|0\rangle&=&if_Pq^{\mu},\\
 	\langle V(q)|\bar{q}\gamma^\mu(1 - \gamma _5)u|0\rangle&=&f_Vm_V \epsilon^{*\mu}
\end{eqnarray}
where $f_P$ and $f_V$ are the decay constants of pseudoscalar and vector mesons, $\epsilon^\mu$ is the polarization vector of vector meson. By substituting Eqs.~(\ref{eq:30}) and (\ref{eq:50}) into Eq.~(\ref{eq:49}), the transition matrices for pseudoscalar and vector mesons are obtained as follows:
\begin{eqnarray}\label{eq:51}
	\notag
	M&=&i\frac{G_F}{\sqrt 2} V_{cb}V^*_{uq}f_Pa_1\\
	\notag
	&&\times\bar u_{\Lambda^*_c[\Xi^*_c]}(p',s') \Bigg\{
		\Bigg[f_1(m_P^2)\slashed{q}+\frac{m_P^2}{m_{\Lambda_b[\Xi_b]}}f_3(m_P^2) \Bigg]\gamma _5\\
	\notag
	&&- \Bigg[g_1(m_P^2)\slashed{q}+\frac{m_P^2}{m_{\Lambda_b[\Xi_b]}}g_3(m_P^2) \Bigg]
	\Bigg\}u_{\Lambda_b[\Xi_b]}(p,s),\\
	\notag
	M&=&\frac{G_F}{\sqrt 2} V_{cb}V^*_{uq}f_Vm_Va_1\\
	\notag
	&&\times\bar u_{\Lambda^*_c[\Xi^*_c]}(p',s')\Bigg\{
		\Bigg[f_1(m_V^2)\slashed{\epsilon}^*+\frac{\slashed{q}\slashed{\epsilon}^*}{m_{\Lambda_b[\Xi_b]}}f_2(m_V^2) \Bigg]\gamma _5\\
	&&- \Bigg[g_1(m_V^2)\slashed{\epsilon}^*+\frac{\slashed{q}\slashed{\epsilon}^*}{m_{\Lambda_b[\Xi_b]}}g_2(m_V^2) \Bigg]
 \Bigg\}u_{\Lambda_b[\Xi_b]}(p',s').
\end{eqnarray}
\section{Numerical results and discussion}\label{sec5}
All input parameters in the present work are collected in Table.~\ref{IP}. The mass of $u(d)$ quark is ignored in this analysis, the masses of $s$, $c$ and $b$ quarks and the values of vacuum condensate are energy scale dependence and fulfill the following renormalization group equations~\cite{ParticleDataGroup:2026mpi},
\begin{eqnarray}\label{eq:52}
	\notag
	m_s(\mu ) &&= m_s(2 \mathrm{GeV})\left[\frac{\alpha _s(\mu )}{\alpha _s(2 \mathrm{GeV})}\right]^{\frac{12}{33 - 2N_f}},\\
	\notag
	m_{c[b]}(\mu ) &&= m_{c[b]}(m_{c[b]})\left[\frac{\alpha _s(\mu )}{\alpha _s(m_{c[b]})}\right]^{\frac{12}{33 - 2N_f}},\\
	\notag
	\left\langle \bar qq \right\rangle (\mu ) &&= \left\langle \bar qq \right\rangle (1\mathrm{GeV})\left[\frac{\alpha _s(1\mathrm{GeV})}{\alpha _s(\mu )}\right]^{\frac{12}{33 - 2N_f}},\\
	\notag
	\left\langle \bar ss \right\rangle (\mu ) &&= \left\langle \bar ss \right\rangle (1\mathrm{GeV})\left[\frac{\alpha _s(1\mathrm{GeV})}{\alpha _s(\mu )}\right]^{\frac{12}{33 - 2N_f}},\\
	\notag
	\left\langle \bar qg_s\sigma Gq \right\rangle (\mu ) &&= \left\langle \bar qg_s\sigma Gq \right\rangle (1\mathrm{GeV})\left[\frac{\alpha _s(1\mathrm{GeV})}{\alpha _s(\mu )}\right]^{\frac{2}{33 - 2N_f}},\\
	\notag
	\left\langle \bar sg_s\sigma Gs \right\rangle (\mu ) &&= \left\langle \bar sg_s\sigma Gs \right\rangle (1\mathrm{GeV})\left[\frac{\alpha _s(1\mathrm{GeV})}{\alpha _s(\mu )}\right]^{\frac{2}{33 - 2N_f}},\\
\end{eqnarray}
with
\begin{eqnarray}\label{eq:53}
	\notag
	\alpha _s(\mu ) &&= \frac{1}{b_0t}\left[ 1 - \frac{b_1}{b_0^2}\frac{\log t}{t} + \frac{b_1^2(\log ^2t - \log t - 1) + b_0b_2}{b_0^4t^2} \right],\\
\end{eqnarray}
where $t=\log\left(\frac{\mu^2}{\Lambda_{\mathrm{QCD}}^2}\right)$, $b_0=\frac{33-2N_f}{12\pi}$, $b_1=\frac{153-19N_f}{24\pi^2}$ and $b_2=\frac{2857-\frac{5033}{9}N_f+\frac{325}{27}N_f^2}{128\pi^3}$. The value of $\Lambda_{\mathrm{QCD}}$ is taken as $0.213$, $0.292$ and $0.343$ GeV for the quark flavors $N_f=5$, $4$ and $3$. The minimum subtraction masses of $s$, $c$ and $b$ quarks are taken from the PDG~\cite{ParticleDataGroup:2026mpi}, which are $m_s(\mu=2 \mathrm{GeV})=0.093\pm0.0007$ GeV, $m_c(m_c)=1.273\pm0.005$ GeV and $m_b(m_b)=4.186\pm0.006$ GeV.
\begin{table}
	\begin{ruledtabular}\caption{Input parameters (IP) in this work. The values of vacuum condensate are at the energy scale $\mu=1$ GeV.}
		\label{IP}
		\renewcommand\arraystretch{1.3}
		\begin{tabular}{c c c c}
			IP&Values&IP&Value \\\hline
			$\langle\bar{q}q\rangle$&$-(0.24\pm0.01)$ GeV$^3$~\cite{Shifman:1978by,Reinders:1984sr}&$m_\rho$&$0.775$ GeV~\cite{ParticleDataGroup:2026mpi} \\
			$\langle\bar{s}s\rangle$&$(0.8\pm0.1)\langle\bar{q}q\rangle$~\cite{Shifman:1978by,Reinders:1984sr}&$m_{K^*}$&$0.892$ GeV~\cite{ParticleDataGroup:2026mpi} \\
			$\langle\bar{q}g_s\sigma Gq\rangle$&$m_0^2\langle\bar{q}q\rangle$~\cite{Shifman:1978by,Reinders:1984sr}&$f_\pi$&$0.131$ GeV~\cite{ParticleDataGroup:2026mpi} \\
			$\langle\bar{s}g_s\sigma Gs\rangle$&$m_0^2\langle\bar{s}s\rangle$~\cite{Shifman:1978by,Reinders:1984sr}&$f_K$&$0.160$ GeV~\cite{ParticleDataGroup:2026mpi} \\
			$m_0^2$&$(0.8\pm0.1)$ GeV$^2$~\cite{Shifman:1978by,Reinders:1984sr}&$f_\rho$&$0.216$ GeV~\cite{ParticleDataGroup:2026mpi} \\
			$\langle g_s^2GG\rangle$&$0.47\pm0.15$ GeV$^4$~\cite{Narison:2010cg,Narison:2011xe,Narison:2011rn}&$f_{K^*}$&$0.217$ GeV~\cite{ParticleDataGroup:2026mpi} \\
			$m_e$&$0.511$ MeV~\cite{ParticleDataGroup:2026mpi}&$V_{cb}$&$0.041$~\cite{ParticleDataGroup:2026mpi} \\
			$m_\mu$&$105.7$ MeV~\cite{ParticleDataGroup:2026mpi}&$V_{ud}$&$0.974$~\cite{ParticleDataGroup:2026mpi} \\
			$m_\tau$&$1.78$ GeV~\cite{ParticleDataGroup:2026mpi}&$V_{us}$&$0.224$~\cite{ParticleDataGroup:2026mpi} \\
			$m_\pi$&$0.140$ GeV~\cite{ParticleDataGroup:2026mpi}&$a_1$&$1.07$~\cite{Buchalla:1995vs} \\
			$m_K$&$0.494$ GeV~\cite{ParticleDataGroup:2026mpi}&~&~ \\
		\end{tabular}
	\end{ruledtabular}
\end{table}
\subsection{The mass and pole residue of $\Lambda$-type single heavy baryon}\label{sec5A}
\begin{figure}
	\centering
	\includegraphics[width=9cm]{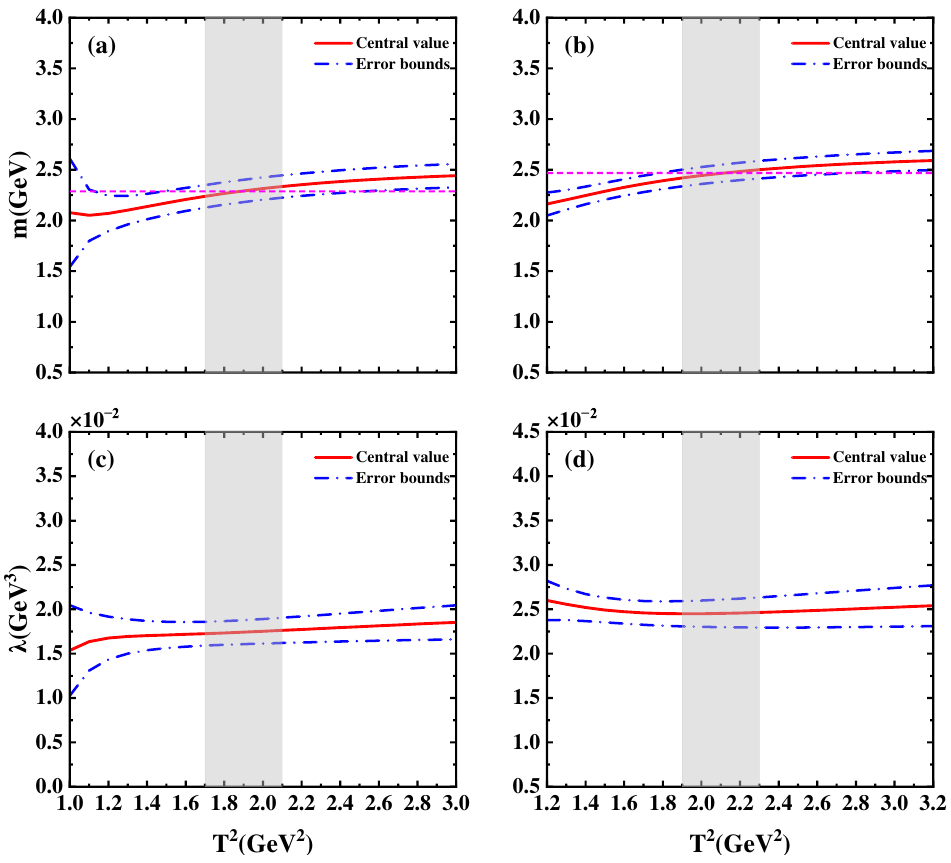}
	\caption{The mass and pole residue for $\Lambda_c$ (a and c) and $\Xi_c$ (b and d) with variation of the Borel parameter $T^2$, where the grey bound denote the Borel platform and the pink dashed line denotes the experimental values.}\label{BWLcXc}
\end{figure}

\begin{figure}
	\centering
	\includegraphics[width=9cm]{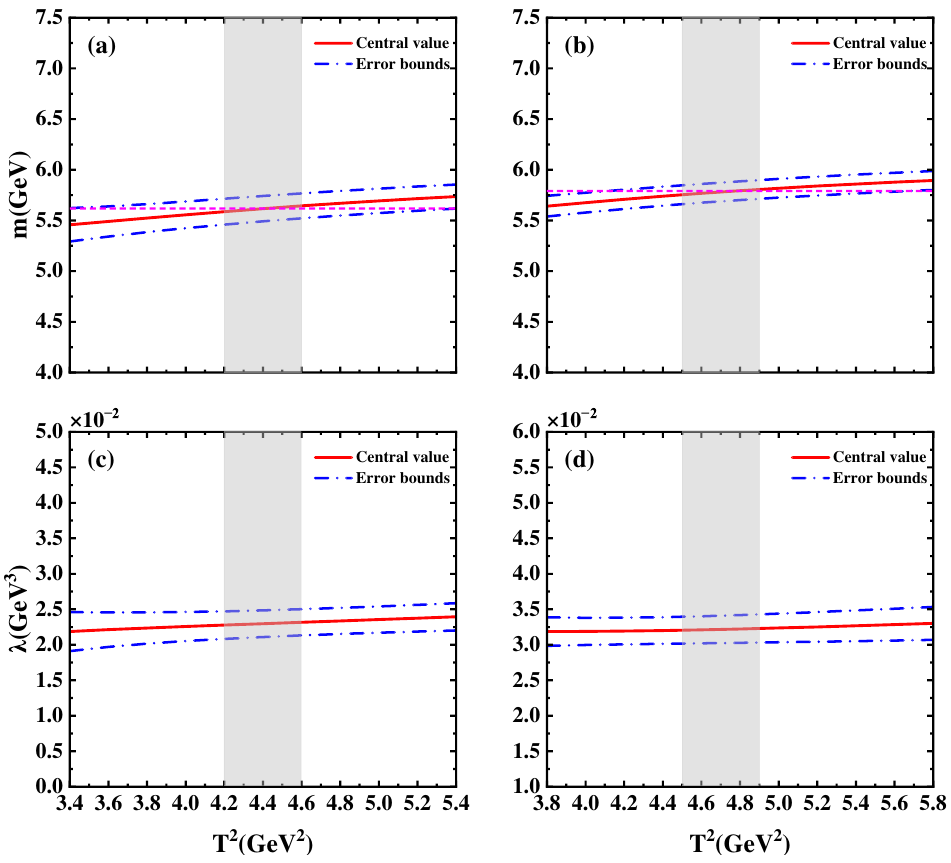}
	\caption{The mass and pole residue for $\Lambda_b$ (a and c) and $\Xi_b$ (b and d) with variation of the Borel parameter $T^2$, where the grey bound denote the Borel platform and the pink dashed line denotes the experimental values.}\label{BWLbXb}
\end{figure}

The masses and pole residues of $\Lambda_{c[b]}$ and $\Xi_{c[b]}$ are determined by Eq.~(\ref{eq:21}). From this equation, the QCD sum rule results of mass and pole residue depend on the quark mass, vacuum condensate, continuum threshold parameter $s_0$ and the Borel parameter $T^2$. Firstly, we need to choose the appropriate energy scale to determine the values of quark masses and vacuum condensates. For $\Lambda_c$ and $\Lambda_b$, the typical energy scale $\mu=1$ and $2$ GeV are adopted~\cite{Wang:2020mxk}, while for $\Xi_c$ and $\Xi_b$, we take $\mu=1.1$ and $2.1$ GeV, where the $0.1$ GeV originates from the difference of $s$ quark. An appropriate work region of Borel parameter should be selected to obtain reliable results. The mass and pole residue should have a weak Borel parameter dependency in this region. This work region is commonly named as ’Borel platform’. At the same time, the pole dominance and the convergence of OPE should be satisfied. The pole contribution and the $n$-dimension vacuum condensate contribution can be defined as,
\begin{eqnarray}\label{eq:54}
	\notag
	\mathrm{Pole}(T^2)&=&\frac{\int_{s_{\min}}^{s_0} dse^{- s/T^2} \left[\sqrt{s}\rho^{1\mathrm{QCD}}_+(s) + \rho^{2\mathrm{QCD}}_+(s)\right]}{\int_{s_{\min}}^{\infty} dse^{- s/T^2}\left[\sqrt{s}\rho^{1\mathrm{QCD}}_+(s) + \rho^{2\mathrm{QCD}}_+(s)\right]},\\
	\notag
	D_n&=&\frac{\int_{s_{\min}}^{s_0} dse^{- s/T^2} \left[\sqrt{s}\rho^{1\mathrm{QCD}(n)}_+(s) + \rho^{2\mathrm{QCD}(n)}_+(s)\right]}{\int_{s_{\min}}^{s_0} dse^{- s/T^2}\left[\sqrt{s}\rho^{1\mathrm{QCD}}_+(s) + \rho^{2\mathrm{QCD}}_+(s)\right]},\\
\end{eqnarray}
here, $\rho^{i\mathrm{QCD}(n)}$ represents the QCD spectral density of $n$ dimension vacuum condensate. Pole dominance requires the pole contribution satisfy $\mathrm{Pole}\approx50\%$, while the convergence of OPE demands that contributions from high dimension vacuum condensate remain small. After repeated trials and comparisons, the Borel platforms for the mass and pole residue of $\Lambda_{c[b]}$ and $\Xi_{c[b]}$ are determined, as shown in Figs.~\ref{BWLcXc} and \ref{BWLbXb}. At the center of these Borel platforms, the pole contribution is larger than 50\%, and the contribution of the dimension-14 and -15 vacuum condensate is well below 1\%. This ensures that both pole dominance and OPE convergence are satisfied.

\begin{figure}
	\centering
	\includegraphics[width=9cm]{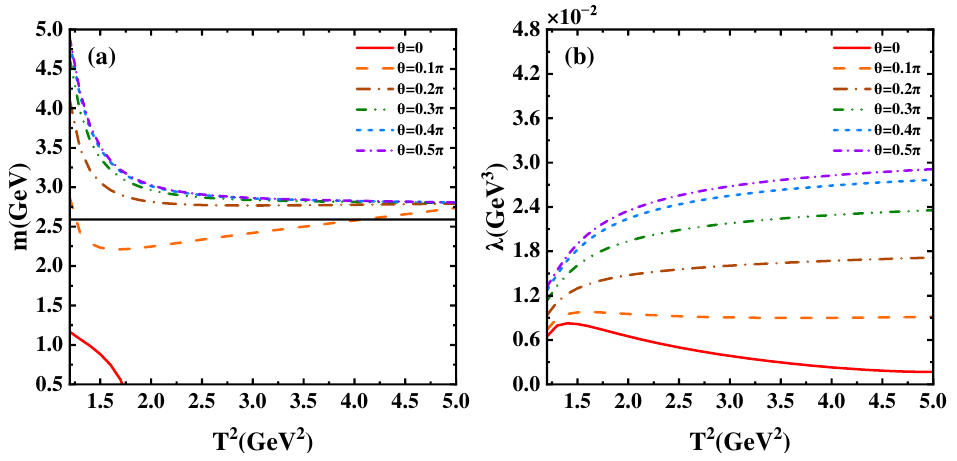}
	\caption{The mass (a) and pole residue (b) for $\Lambda^*_c(2595)$ with variation of the Borel parameter $T^2$ in different mixing angle $\theta$, where the black real line denotes the experimental values. The continuum threshold parameter is taken as $\sqrt{s_0}=3.10$ GeV.}\label{theta}
\end{figure}

\begin{figure}
	\centering
	\includegraphics[width=9cm]{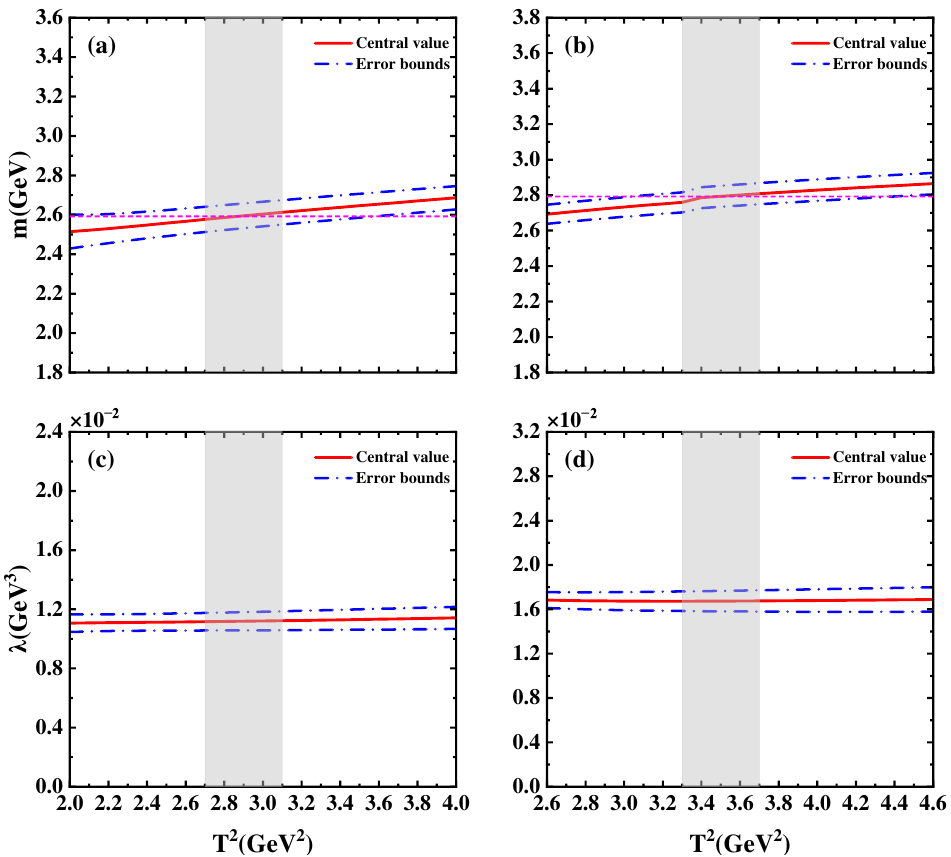}
	\caption{The mass and pole residue for $\Lambda^*_c$ (a and c) and $\Xi^*_c$ (b and d) with variation of the Borel parameter $T^2$, where the grey bound denote the Borel platform and the pink dashed line denotes the experimental values.}\label{BWLcsXcs}
\end{figure}

\begin{figure}
	\centering
	\includegraphics[width=9cm]{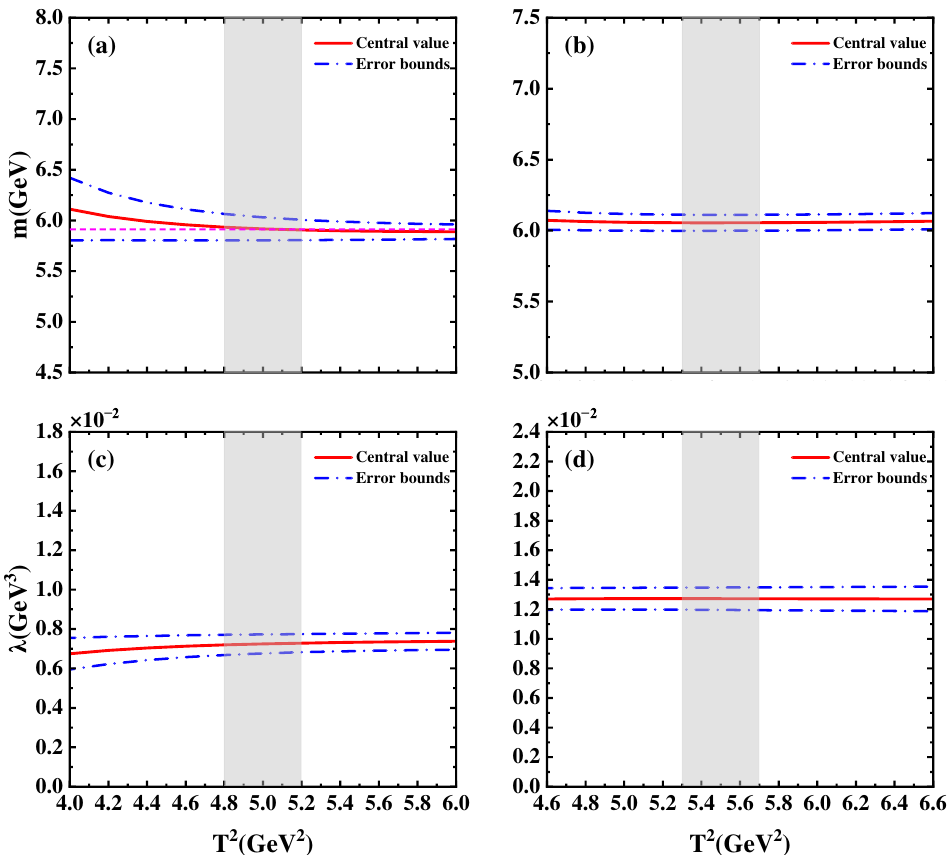}
	\caption{The mass and pole residue for $\Lambda^*_b$ (a and c) and $\Xi^*_b$ (b and d) with variation of the Borel parameter $T^2$, where the grey bound denote the Borel platform and the pink dashed line denotes the experimental values.}\label{BWLbsXbs}
\end{figure}

The masses and pole residues of $\Lambda^*_{c[b]}$ and $\Xi^*_{c[b]}$ are determined by Eq.~(\ref{eq:26}). Similar to the previous procedures, we first assume that the orbital excitation energy is approximately 0.3 GeV, based on which the energy scales of $\Lambda^*_c$ and $\Xi^*_c$ are determined to be 1.3 and 1.4 GeV, and those of $\Lambda^*_b$ and $\Xi^*_b$ are 2.3 and 2.4 GeV. The pole contribution and the $n$-dimension vacuum condensate contribution for negative parity baryon can be defined as,
\begin{eqnarray}\label{eq:55}
	\notag
	\mathrm{Pole}(T^2)&=&\frac{\int_{s_{\min}}^{s_0} dse^{-s/T^2} \left[m_{\Lambda_Q[\Xi_Q]}\rho^{1\mathrm{QCD}}_-(s) + \rho^{2\mathrm{QCD}}_-(s)\right]}{\int_{s_{\min}}^{\infty} dse^{-s/T^2} \left[m_{\Lambda_Q[\Xi_Q]}\rho^{1\mathrm{QCD}}_-(s) + \rho^{2\mathrm{QCD}}_-(s)\right]},\\
	\notag
	D_n&=&\frac{\int_{s_{\min}}^{s_0} dse^{-s/T^2} \left[m_{\Lambda_Q[\Xi_Q]}\rho^{1\mathrm{QCD}(n)}_-(s) + \rho^{2\mathrm{QCD}(n)}_-(s)\right]}{\int_{s_{\min}}^{s_0} dse^{-s/T^2} \left[m_{\Lambda_Q[\Xi_Q]}\rho^{1\mathrm{QCD}}_-(s) + \rho^{2\mathrm{QCD}}_-(s)\right]}.\\
\end{eqnarray}
We first discuss the mixing angle by taking $\Lambda_c^*(2595)$ as an example. When the continuum threshold parameter $\sqrt{s_0}$ is taken as $3.10$ GeV (the energy gap $\delta$ is about $0.51$ GeV), we plot the mass and pole residue as functions of the Borel parameter $T^2$ for $\theta$ varying from $0$ to $0.5\pi$ in steps of $0.1\pi$, as shown in Fig.~\ref{theta}. From Fig.~\ref{theta}(a), it is found that for very small mixing angles no stable Borel platform for the mass can be obtained, which implies that the continuum contamination cannot be well suppressed by the pure $J_1$ current. When $\theta$ exceeds $0.2\pi$, although a platform appears, one must lower $\sqrt{s_0}$ to $2.7-2.8$ GeV to reproduce the experimental mass. Such a small threshold parameter corresponds to an energy gap of only $0.1-0.2$ GeV, which is physically unreasonable as it severely violates the pole dominance. We thus restrict the optimal mixing angle to the range $0.1- 0.2\pi$, where the pole residue also exhibits a good platform (see Fig.~\ref{theta}(b)). After repeated trials and adjustments, we finally take $\theta = 0.130\pi$. This indicates that, within the QCD sum rules, the optimal interpolating current for isolating the $\Lambda_c^*(2595)$ state is dominated by the $J_1$ Dirac structure, with a small admixture of $J_2$. Using the determined mixing angle, we then show the Borel platforms for the mass and pole residue of $\Lambda^*_{c[b]}$ and $\Xi^*_{c[b]}$ in Figs.~\ref{BWLcsXcs} and \ref{BWLbsXbs}.


Table~\ref{BP} summarizes the mixing angle for negative-parity states, together with the Borel platform, continuum threshold parameter, energy scale, and pole contribution for the positive- and negative-parity $\Lambda_{c[b]}$ and $\Xi_{c[b]}$ baryons, where the final masses and pole residues are also listed. The uncertainties in the masses and pole residues are estimated by considering the individual effects of the continuum threshold parameter $s_0$, the vacuum condensate parameters, and the quark masses. For each source, we vary the corresponding parameter within its uncertainty range and calculate its impact on the Borel curves of the mass and pole residue. The combined uncertainties from all sources are then obtained taking the square root of the sum of the squared uncertainties, which yields the upper and lower bounds of the Borel curves. Finally, using the upper and lower bounds of the Borel platform, we determine the uncertainties of the final masses and pole residues, as listed in Table~\ref{BP}.
\begin{table*}
	\begin{ruledtabular}\caption{The mixing angle (only for negative-parity baryon), Borel platform, continuum threshold parameter, energy scale, pole contribution, mass and pole residue for all $\Lambda$-type singly heavy baryons.}\label{BP}
		\label{mandlambda}
		\renewcommand\arraystretch{1.3}
		\begin{tabular}{c c c c c c c c c}
			Baryon&$\theta$&$T^2$ (GeV$^2$)&$\sqrt{s_0}$ (GeV)&$\mu$ (GeV)&Pole(\%)&$m$ (GeV)&$\lambda$ (GeV$^3$)&Experiment~\cite{ParticleDataGroup:2026mpi} \\\hline
			$\Lambda_c$&$-$&$1.7-2.1$&$2.92\pm0.10$&$1.0$&$60-42$&$2.29\pm0.17$&$0.0174\pm0.0015$&$2.286$ \\
			$\Xi_c$&$-$&$1.9-2.3$&$3.05\pm0.10$&$1.1$&$60-45$&$2.46\pm0.12$&$0.0245\pm0.0016$&$2.468$ \\
			$\Lambda_b$&$-$&$4.2-4.6$&$6.41\pm0.10$&$2.0$&$55-46$&$5.62\pm0.16$&$0.0230\pm0.0020$&$5.620$ \\
			$\Xi_b$&$-$&$4.5-4.9$&$6.51\pm0.10$&$2.1$&$55-47$&$5.79\pm0.12$&$0.0322\pm0.0020$&$5.792$ \\
			$\Lambda^*_c$&$0.130\pi$&$2.7-3.1$&$3.10\pm0.10$&$1.3$&$64-58$&$2.60\pm0.08$&$0.0112\pm0.0007$&$2.592$ \\
			$\Xi^*_c$&$0.135\pi$&$3.3-3.7$&$3.30\pm0.10$&$1.4$&$54-48$&$2.79\pm0.09$&$0.0167\pm0.0010$&$2.792$ \\
			$\Lambda^*_b$&$0.085\pi$&$4.8-5.2$&$6.35\pm0.10$&$2.3$&$74-63$&$5.92\pm0.10$&$0.0072\pm0.0005$&$5.912$ \\
			$\Xi^*_b$&$0.100\pi$&$5.3-5.7$&$6.48\pm0.10$&$2.4$&$75-66$&$6.05\pm0.05$&$0.0127\pm0.0008$&$6.087?$ \\
		\end{tabular}
	\end{ruledtabular}
\end{table*}

\subsection{Numerical results of form factor}\label{sec5B}
The form factors of $\Lambda_b\to\Lambda^*_c$ and $\Xi_b\to\Xi^*_c$ are determined by Eq.~(\ref{eq:38}). From this equation, we can find that the results of QCD sum rules for form factors depend on several parameters such as the Borel parameter $T^2$, continuum threshold parameters $s_0$ and $u_0$, and the square of transition momentum $Q^2$. For $\Lambda_b$ and $\Xi_b$ decays, we take the typical energy scales $\mu=2$ and $2.1$ GeV. The values of continuum threshold parameters are determined by the calculations of two-point QCDSR in Sec.~\ref{sec5A}. As with the two-point QCDSR, we first need to determine the 'Borel platform' for form factors based on the pole dominance and the convergence of the OPE. To discuss the pole dominance, the following definitions for pole contributions of $s$ and $u$ channels are given as,
\begin{eqnarray}\label{eq:56}
	\mathrm{Pole}_s=\frac{\int_{u_{\min }}^{u_0} du \int_{s_{\min }}^{s_0} ds}{\int_{u_{\min }}^{u_0} du \int_{s_{\min }}^{\infty} ds} ,
	\mathrm{Pole}_u=\frac{\int_{u_{\min }}^{u_0} du \int_{s_{\min }}^{s_0} ds}{\int_{u_{\min }}^{\infty} du \int_{s_{\min }}^{s_0} ds}.
\end{eqnarray}
The pole dominance requires that the pole contributions for $s$ and $u$ channels should be both larger than 40\% in the present work.

\begin{figure*}
	\centering
	\includegraphics[width=17cm]{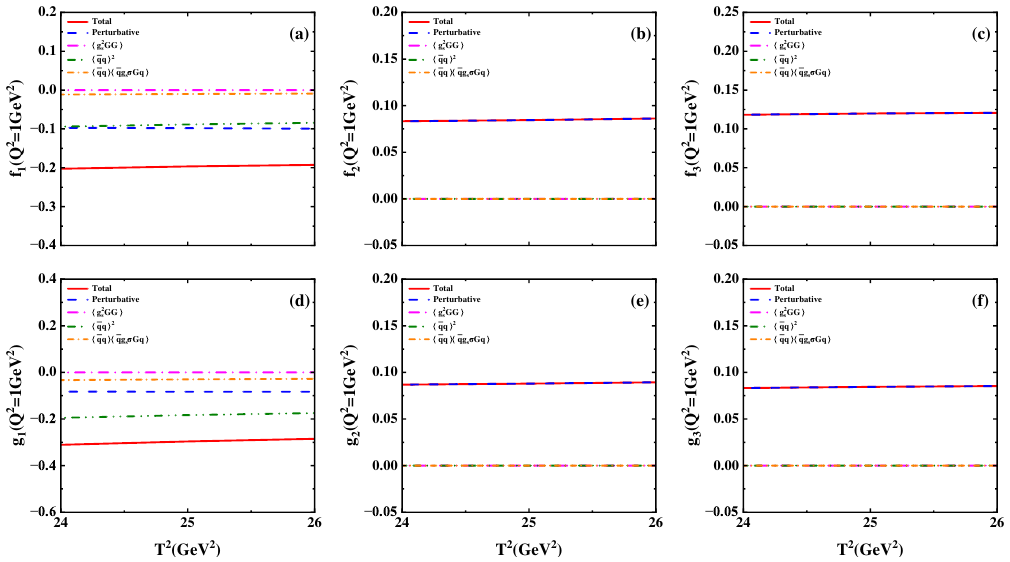}
	\caption{The contributions of the perturbative part and different vacuum condensate terms with variation of Borel parameter $T^2$ for the $\Lambda_b\to\Lambda^*_c$ transition form factors.}\label{BWFFLbtoLcs}
\end{figure*}

\begin{figure*}
	\centering
	\includegraphics[width=17cm]{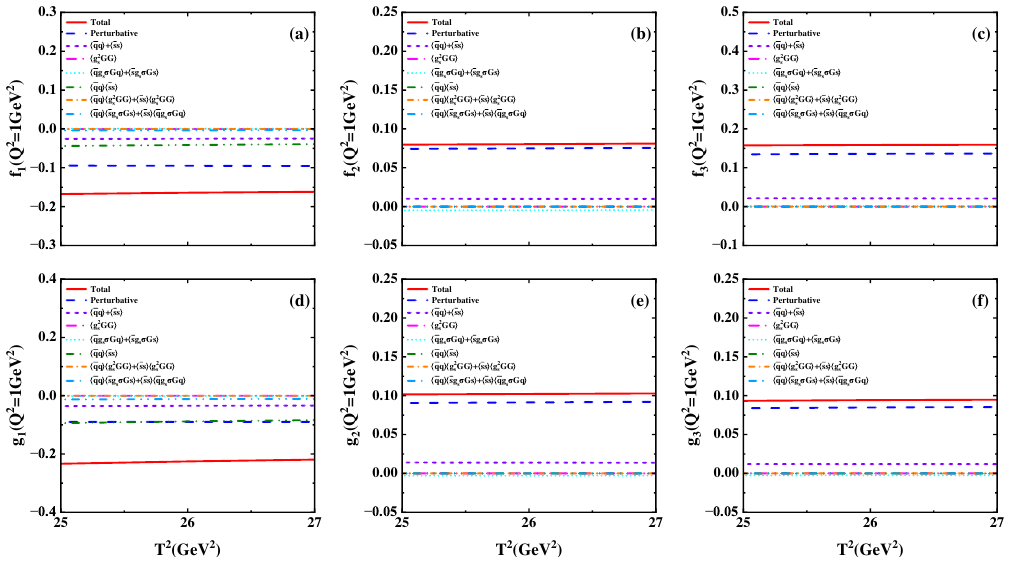}
	\caption{The contributions of the perturbative part and different vacuum condensate terms with variation of Borel parameter $T^2$ for the $\Xi_b\to\Xi^*_c$ transition form factors.}\label{BWFFXbtoXcs}
\end{figure*}

With $Q^2$ fixed at 1 GeV$^2$, we determine the Borel platforms for all form factors to be 24–26 GeV$^2$ for $\Lambda_b\to\Lambda^*_c$ transition and 25–27 GeV$^2$ for $\Xi_b\to\Xi^*_c$ transition through repeated trials and comparisons. These results are presented in Figs.~\ref{BWFFLbtoLcs} and \ref{BWFFXbtoXcs}. Besides, the total contribution, the perturbative part and the contributions from different vacuum condensates are also show in Figs.~\ref{BWFFLbtoLcs} and \ref{BWFFXbtoXcs}. Firstly, the trend of all form factors with the Borel parameter is relatively flat in determined Borel platforms. For $\Lambda_b\to\Lambda^*_c$ transition, the contributions from vacuum condensates $\langle\bar{q}q\rangle$, $\langle\bar{q}g_s\sigma Gq\rangle$ and $\langle\bar{q}q\rangle\langle g_s^2GG\rangle$ can be neglected because they are proportional to the mass of $u(d)$ quark. From Fig.~\ref{BWFFLbtoLcs}, one can find that the contribution from gloun condensate $\langle g_s^2GG\rangle$ term is less than 1\%, the main contributions for form factors $f_1$ and $g_1$ come from the perturbation part and the four quark condensate $\langle\bar{q}q\rangle^2$. Specifically, for $f_1$, the contributions from each dimension follow the order $pert\approx\langle\bar{q}q\rangle^2>>\langle\bar{q}q\rangle\langle\bar{q}g_s\sigma Gq\rangle$, while for $g_1$, the order is $\langle\bar{q}q\rangle^2>pert>>\langle\bar{q}q\rangle\langle\bar{q}g_s\sigma Gq\rangle$. The contribution of the 8-dimension term $\langle\bar{q}q\rangle\langle\bar{q}g_s\sigma Gq\rangle$ is less than 8\%. For the form factors $f_{2,3}$ and $g_{2,3}$, except for the perturbative part, the contributions of the remaining vacuum condensate terms all tend to zero. For $\Xi_b\to\Xi^*_c$ transition (See Fig.~\ref{BWFFXbtoXcs}), the main contributions for form factors $f_1$ and $g_1$ come from the perturbation part, quark condensate $\langle\bar{q}q\rangle$ or $\langle\bar{s}s\rangle$ and four quark condensate $\langle\bar{q}q\rangle\langle\bar{s}s\rangle$. For $f_1$, the contributions follow the order $pert>\langle\bar{q}q\rangle\langle\bar{s}s\rangle>\langle\bar{q}q\rangle+\langle\bar{s}s\rangle>>\langle\bar{q}q\rangle\langle\bar{s}g_s\sigma Gs\rangle+\langle\bar{s}s\rangle\langle\bar{q}g_s\sigma Gq\rangle$, and for $g_1$, the order is $pert\approx\langle\bar{q}q\rangle\langle\bar{s}s\rangle>\langle\bar{q}q\rangle+\langle\bar{s}s\rangle>>\langle\bar{q}q\rangle\langle\bar{s}g_s\sigma Gs\rangle+\langle\bar{s}s\rangle\langle\bar{q}g_s\sigma Gq\rangle$. The contribution of the 8-dimension term $\langle\bar{q}q\rangle\langle\bar{s}g_s\sigma Gs\rangle$ and $\langle\bar{s}s\rangle\langle\bar{q}g_s\sigma Gq\rangle$ is less than 2\%. For the form factors $f_{2,3}$ and $g_{2,3}$, the main contributions come from perturbative part and quark condensate $\langle\bar{q}q\rangle+\langle\bar{s}s\rangle$, the contribution of quark condensate is less than 10\%. These all demonstrate good OPE convergence.

The uncertainty analysis for the form factor is of great importance. According to Eq.~(\ref{eq:38}), the uncertainties of the form factor originate from numerous input parameters,
these input parameters can be divided into two types: one at hardon level, such as the
pole residues of baryons, and the other at QCD level, such as the vacuum condensates, quark mass, threshold parameters and Borel parameter. In the present work, we neglect the uncertainty caused by the mass of hadrons. To estimate the uncertainty of form factor, we uniformly write
all QCD sum rule equations in Eq.~(\ref{eq:38}) in the following form:
\begin{eqnarray}\label{eq:57}
	\lambda_{\mathcal{B}_i}\lambda_{\mathcal{B}_f}F=f(x_i),
\end{eqnarray}
where $\lambda_{\mathcal{B}_i}$ and $\lambda_{\mathcal{B}_f}$ are pole residues of initial and final state baryons, $F$ denote the form factor and $f(x_i)$ is a numerical integral associated with QCD level physical quantities, with $x_i$ denotes the input parameters,
\begin{eqnarray}
	\notag
	x_i&=&\Big\{\langle\bar{q}q\rangle, \langle\bar{s}s\rangle, \langle\bar{q}g_s\sigma Gq\rangle, \langle\bar{s}g_s\sigma Gs\rangle, \langle g_s^2GG\rangle,\\
	\notag
	&& m_s, m_c, m_b, s_0,u_0, T^2\Big\}
\end{eqnarray}
The left and right sides of Eq.~(\ref{eq:57}) are the quantities at the hadron and QCD levels, respectively. By taking the partial derivative of both sides of Eq.~(\ref{eq:57}), we can get:
\begin{eqnarray}
	\frac{\delta\lambda_{\mathcal{B}_i}}{\bar{\lambda}_{\mathcal{B}_i}}+\frac{\delta\lambda_{\mathcal{B}_f}}{\bar{\lambda}_{\mathcal{B}_f}}+\frac{\delta F}{\bar{F}}=\frac{\delta f(x_i)}{\bar{\lambda}_{\mathcal{B}_i}\bar{\lambda}_{\mathcal{B}_f}\bar{F}},
\end{eqnarray}
where $\bar{f}$ denote the central value of quantity $f$. It is worth noting that the threshold parameters and pole residues used are obtained from two point QCD sum rule calculations, and their uncertainties are closely correlated. If we consider their contributions simultaneously, the uncertainty in the form factor would be excessively amplified. Therefore, For simplicity and
also to avoid over estimating uncertainties, the following approximate relation is taken,
\begin{eqnarray}
	\frac{\delta\lambda_{\mathcal{B}_i}}{\bar{\lambda}_{\mathcal{B}_i}}\approx\frac{\delta\lambda_{\mathcal{B}_f}}{\bar{\lambda}_{\mathcal{B}_f}}\approx\frac{\delta F}{\bar{F}},
\end{eqnarray}
Then, the uncertainty for form factor can be expressed as:
\begin{eqnarray}
	\delta F=\frac{\delta f(x_i)}{3\bar{\lambda}_{\mathcal{B}_i}\bar{\lambda}_{\mathcal{B}_f}},
\end{eqnarray}
here, $f(x_i)$ is a numerical integral of the QCD level input, its partial derivative cannot be obtained directly. Therefore, we introduce the following standard deviation formula:
\begin{eqnarray}
	\delta f(x_i)=\sqrt{\sum_i \left[f(\bar{x}_i+\delta x_i)-f(\bar{x}_i)\right]^2}.
\end{eqnarray}
Finally, the  uncertainty for the form factor can be written as:
\begin{eqnarray}
	\delta F=\frac{\sqrt{\sum_i \left[f(\bar{x}_i+\delta x_i)-f(\bar{x}_i)\right]^2}}{3\bar{\lambda}_{\mathcal{B}_i}\bar{\lambda}_{\mathcal{B}_f}},
\end{eqnarray}

\begin{figure*}
	\centering
	\includegraphics[width=17cm]{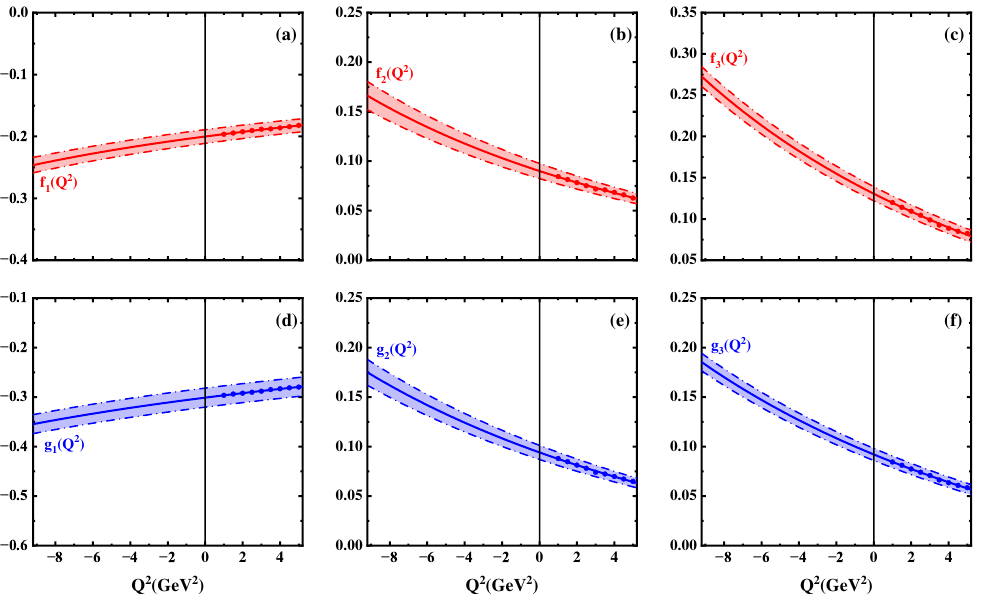}
	\caption{The fitting results of vector (a-c) and axial-vector (d-f) form factors for $\Lambda_b\to\Lambda^*_c$ transition, where the red and blue real lines are fitting curves of central values for vector and axial-vector form factors and the red and blue dash lines are fitting curves of upper and lower bounds.}\label{FFL}
\end{figure*}

\begin{figure*}
	\centering
	\includegraphics[width=17cm]{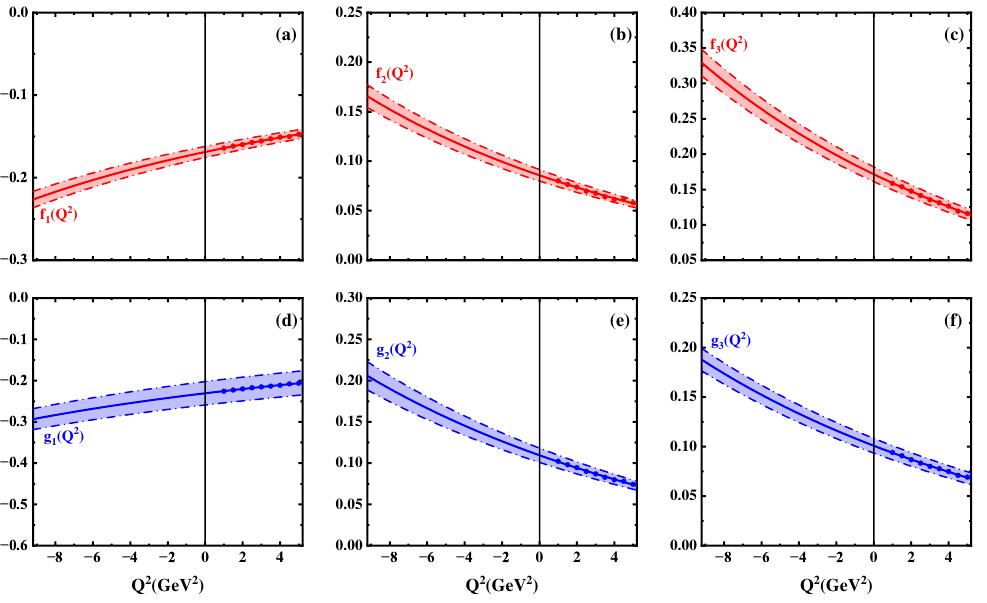}
	\caption{It is the same as Fig.~\ref{FFL}, but for $\Xi_b\to\Xi^*_c$ transition.}\label{FFX}
\end{figure*}

By varying $Q^2$ in the range $1-5$ GeV$^2$, the form factors in the space-like region are obtained. Their values in the time-like region were then determined by fitting the space-like results with an appropriate analytical function and extrapolating them to the time-like region. In this work, the $z$-series expansion approach is employed to fit the form factors because of its high analytical, convergence, and low model dependence~\cite{Boyd:1994tt}. In the present work, the following simplified form of the $z$-series expansion is employed:
\begin{eqnarray}
	\notag
	F(Q^2)&=&\frac{F(0)}{1+Q^2/m_R^2}\\
	\notag
	&&\times\left\{1+a\left[z(Q^2)-z(0)+\frac{1}{2}[z(Q^2)^2-z(0)^2]\right]\right\},\\
\end{eqnarray}
where $m_R=m_{B^*_c(1^-)}(m_{B^*_c(1^+)})$ is for vector (axial-vector) form factor. The masses of ground state vector and axial-vector $B_c$ meson are taken from the results of the Godfrey-Isgur quark model, which are $m_{B^*_c(1^-)}=6.338$ GeV and $m_{B^*_c(1^+)}=6.745$ GeV~\cite{Li:2023wgq}. $F(0)$ and $a$ are the fitting parameters and the $z(Q^2)$ function is taken as the following from,
\begin{eqnarray}
 	z(Q^2)=\frac{\sqrt{t_+^2+Q^2}-\sqrt{t_+^2-t_-^2}}{\sqrt{t_+^2+Q^2}+\sqrt{t_+^2-t_-^2}},
\end{eqnarray}
where $t_\pm^2=(m_{\Lambda_b[\Xi_b]}\pm m_{\Lambda^*_c[\Xi^*_c]})^2$. Since the form factor errors at different $Q^2$ points originate from the same set of QCD input parameters, they are strongly correlated. Treating them as independent in a $\chi^2$ fit would therefore misrepresent the actual error propagation. A fit using the full covariance matrix is in principle possible, but the corresponding correlation matrix is nearly singular and requires a reliable estimate of the correlations among the input parameters, which is not available in practice. We therefore fit the central, upper, and lower bounds of the space-like form factors separately and extrapolate them to the physical region. The resulting uncertainties in the time-like region should be regarded as a conservative envelope estimate rather than a statistical confidence interval. For simplicity, we only present the fitting results of the central values of the form factors in the article, as shown in Table~\ref{ZPC}, with the corresponding fitting images shown in Figs.~\ref{FFL} and \ref{FFX}.
\begin{table}[htbp]
\begin{ruledtabular}\caption{Fitting parameters for the central values of form factors using the $z$-series expansion, and the form factor values at $q^2= q_{\max}^2=(m_{\Lambda_b[\Xi_b]}-m_{\Lambda^*_c[\Xi^*_c]})^2$.}\label{ZPC}
\renewcommand\arraystretch{1.3}
\begin{tabular}{c c c c c}
Mode&Form factor&$F(0)$&$a$&$F(q^2_{\max})$\\ \hline
\multirow{6}*{$\Lambda_b \to \Lambda^*_c$}&$f_1$ &$-0.200$&$1.359$&$-0.246$ \\
~&$f_2$ &$0.0901$&$-11.316$&$0.166$ \\
~&$f_3$ &$0.131$&$-16.333$&$0.272$\\
~&$g_1$&$-0.301$&$1.618$&$-0.354$ \\
~&$g_2$&$0.0941$&$-12.944$&$0.174$\\
~&$g_3$&$0.0921$&$-16.226$&$0.185$\\ \hline
\multirow{6}*{$\Xi_b \to \Xi^*_c$}&$f_1$ &$-0.169$&$-1.026$&$-0.225$ \\
~&$f_2$ &$0.0858$&$-14.449$&$0.163$ \\
~&$f_3$ &$0.172$&$-14.183$&$0.325$ \\
~&$g_1$&$-0.231$&$-0.428$&$-0.292$\\
~&$g_2$&$0.110$&$-14.730$&$0.203$\\
~&$g_3$&$0.101$&$-14.343$&$0.185$\\
\end{tabular}
\end{ruledtabular}
\end{table}

\subsection{Numerical results of semileptonic and nonleptonic decays}\label{sec5C}
\begin{figure*}
	\centering
	\includegraphics[width=17cm]{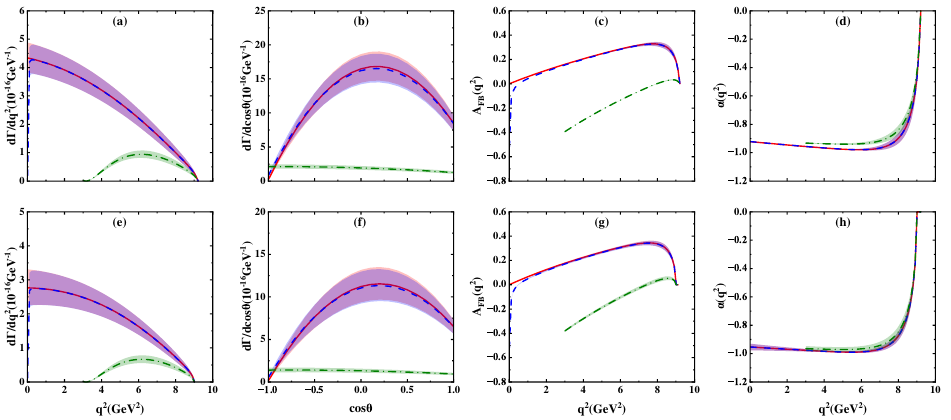}
	\caption{The differential decay width $d\Gamma/dq^2$ (a, e), leptonic forward-backward asymmetry $A_{FB}$ (c, g) and asymmetry parameter $\alpha$ (d, h) with the variables of $q^2$ and the angle distribution $d\Gamma/d\cos\theta_\ell$ (b, f) with the variables of $\cos\theta_\ell$ for semileptonic decay processes $\Lambda_b\to\Lambda^*_c\ell^-\bar{\nu}_\ell$ (a, b, c and d) and $\Xi_b\to\Xi^*_c\ell^-\bar{\nu}_\ell$ (e, f, g and h), where the red, blue and green represent $\ell=e,\mu$ and $\tau$, respectively.}\label{DWLX}
\end{figure*}

With the obtained form factors, the semileptonic decay processes $\Lambda_b\to\Lambda^*_c\ell\bar{\nu}_\ell$ and $\Xi_b\to\Xi^*_c\ell\bar{\nu}_\ell$ can be analyzed. Firstly, integrating Eq.~(\ref{eq:46}) with respect to $\cos\theta_\ell$ and $q^2$ respectively yields the differential decay width and angular distribution as follows:
\begin{eqnarray}\label{eq:65}
	\notag
	\frac{d\Gamma}{dq^2}&=&\int_{-1}^{1}d\cos\theta_\ell\frac{d^2\Gamma}{dq^2d\cos\theta_\ell},\\
	\frac{d\Gamma}{d\cos\theta_\ell}&=&\int_{m_\ell^2}^{(m_{\Lambda_b[\Xi_b]}-m_{\Lambda^*_c[\Xi^*_c]})^2}dq^2\frac{d^2\Gamma}{dq^2d\cos\theta_\ell},
\end{eqnarray}
the differential decay widths with variations of $q^2$ are shown in Figs.~\ref{DWLX}(a) and (e) and the angular distributions are shown in Figs.~\ref{DWLX}(b) and (f). The total decay width can be obtained by integrating out $q^2$ or $\cos\theta_\ell$ in Eq.~(\ref{eq:65}). The numerical results for the decay widths and branching ratios of the semileptonic decays, together with the corresponding experimental data and other theoretical predictions, are collected in Table~\ref{DWBRSM}. The uncertainties in the decay widths originate from the form factors. For the decay $\Lambda_b\to\Lambda^*_c\ell\bar{\nu}_\ell$ ($\ell=e$ or $\mu$), the PDG reports a branching ratio of $(7.9^{+4.0}_{-3.5})\times10^{-3}$~\cite{ParticleDataGroup:2026mpi}, which is comparable to our prediction. However, since $\Lambda^*_c(2595)$ is difficult to reconstruct, the experimental uncertainties remain large, calling for more precise measurements in the future. In Ref.~\cite{Duan:2024lnw}, the same semileptonic decay was studied using LCSR. Their results depend strongly on the choice of interpolating currents for $\Lambda^*_c$ and on the model adopted for the light-cone distribution amplitudes of the $\Lambda_b$ baryon; therefore, only the maximum and minimum values of their branching ratios are quoted in the table. Our central value is smaller than those obtained from the covariant confined quark model (CCQM)~\cite{Gutsche:2018nks} and the light-front quark model (LFQM)~\cite{Li:2021qod}, yet it remains compatible with the former once the error bars are taken into account. For the semileptonic decay $\Xi_b\to\Xi^*_c\ell\bar{\nu}_\ell$, both theoretical and experimental information is still scarce, and further experimental verification is needed.
\begin{table*}[htbp]
	\renewcommand\arraystretch{1.3}
	\begin{ruledtabular}
		\caption{Decay widths (in $10^{-15}$ GeV) and Branching ratios (in $10^{-3}$) of $\Lambda_{b}$ and $\Xi_{b}$ semileptonic decays. Branching ratios are calculated at $\tau_{\Lambda_b^0}=(1.468\pm0.009)\times10^{-12}$ s and $\tau_{\Xi_b^0}=(1.477\pm0.032)\times10^{-12}$ s~\cite{ParticleDataGroup:2026mpi}.}
\begin{tabular}{c >{\centering}p{6.5em}| c | c c c c c c}\label{DWBRSM}
	&\multirow{2}{*}{Decay channels} & Decay widths & \multicolumn{6}{c}{Branching ratios}\\
	&&This work &This work&Experiment~\cite{ParticleDataGroup:2026mpi}&LCSR~\cite{Duan:2024lnw}&LCSR~\cite{Duan:2024lnw}&CCQM~\cite{Gutsche:2018nks}&LFQM~\cite{Li:2021qod}\\
	\hline
	&$\Lambda_b \to \Lambda^*_{c}e^-\bar{\nu}_{e}$ &$2.488^{+0.333}_{-0.310}$&$5.548^{+0.745}_{-0.691}$ &$7.9^{+4.0}_{-3.5}$&$23.875^{+12.101}_{-8.647}$&$0.079^{+0.084}_{-0.043}$&$8.6\pm1.7$&$17.3\pm5.9$ \\
	&$\Lambda_b \to \Lambda^*_{c}\mu^-\bar{\nu}_{\mu}$ &$2.471^{+0.332}_{-0.307}$&$5.512^{+0.740}_{-0.687}$ &$7.9^{+4.0}_{-3.5}$&$23.735^{+12.024}_{-8.953}$&$0.078^{+0.084}_{-0.042}$&$8.5\pm1.7$&$17.2\pm5.8$ \\
	&$\Lambda_b \to \Lambda^*_{c}\tau^-\bar{\nu}_{\tau}$ &$0.369^{+0.057}_{-0.052}$&$0.823^{+0.127}_{-0.116}$ &$-$&$3.903^{+1.824}_{-1.316}$&$0.015^{+0.016}_{-0.008}$&$1.1\pm0.2$&$2.4\pm1.1$ \\
	&$\Xi_b \to \Xi^*_{c}e^-\bar{\nu}_e$ &$1.717^{+0.292}_{-0.274}$&$3.853^{+0.654}_{-0.614}$ &$-$&$-$&$-$&$-$&$-$ \\
	&$\Xi_b \to \Xi^*_{c}\mu^-\bar{\nu}_\mu$ &$1.706^{+0.290}_{-0.272}$&$3.829^{+0.650}_{-0.611}$ &$-$&$-$&$-$&$-$&$-$ \\
	&$\Xi_b \to \Xi^*_{c}\tau^-\bar{\nu}_\tau$ &$0.254^{+0.048}_{-0.043}$&$0.571^{+0.106}_{-0.098}$ &$-$&$-$&$-$&$-$&$-$ \\
\end{tabular}
	\end{ruledtabular}
\end{table*}

The lepton universality ratio is an important observable for testing the Standard Model. Our predictions for these ratios, together with those from other collaborations, are summarized in Table~\ref{LUR}. As shown in Table~\ref{LUR}, our result for $R_{\Lambda^*_c}$ is close to the predictions of the CCQM~\cite{Gutsche:2018nks} and LFQM~\cite{Li:2021qod}, and also to the lower bound obtained from LCSR~\cite{Duan:2024lnw}. 
\begin{table}[htbp]
	\renewcommand\arraystretch{1.4}
	\begin{ruledtabular}
		\caption{The lepton universality ratios (LUR) of $\Lambda_b$ and $\Xi_b$ semileptonic decays.}
		\begin{tabular}{c c c}\label{LUR}
			LUR&$R_{\Lambda^*_c}=\frac{\mathcal{B}[\Lambda_b\to\Lambda^*_c\tau\bar{\nu}_\tau]}{\mathcal{B}[\Lambda_b\to\Lambda^*_c\mu\bar{\nu}_\mu]}$&$R_{\Xi^*_c}=\frac{\mathcal{B}[\Xi_b\to\Xi^*_c\tau\bar{\nu}_\tau]}{\mathcal{B}[\Xi_b\to\Xi^*_c\mu\bar{\nu}_\mu]}$ \\ \hline
			This work&$0.149^{+0.031}_{-0.028}$&$0.149^{+0.038}_{-0.035}$ \\
			\multirow{2}{*}{LCSR~\cite{Duan:2024lnw}}&$0.285^{+0.006}_{-0.005}$&$-$ \\
			~&$0.164^{+0.006}_{-0.004}$&$-$ \\
			CCQM~\cite{Gutsche:2018nks}&$0.13\pm0.03$&$-$ \\
			LFQM~\cite{Li:2021qod}&$0.14\pm0.01$&$-$ \\
		\end{tabular}
	\end{ruledtabular}
\end{table}
In addition, we also analyze two asymmetry observables which are the leptonic forward-backward asymmetry $A_{FB}$ and asymmetry parameter $\alpha$. The definitions of those two asymmetry observables are as follows:
\begin{eqnarray}\label{eq:66}
	\notag
	A_{FB}(q^2)&=&\frac{\int_{0}^{1}d\cos\theta_\ell\frac{d^2\Gamma}{dq^2d\cos\theta_\ell}-\int_{-1}^{0}d\cos\theta_\ell\frac{d^2\Gamma}{dq^2d\cos\theta_\ell}}{\int_{0}^{1}d\cos\theta_\ell\frac{d^2\Gamma}{dq^2d\cos\theta_\ell}+\int_{-1}^{0}d\cos\theta_\ell\frac{d^2\Gamma}{dq^2d\cos\theta_\ell}},\\
	\alpha(q^2)&=&\frac{d\Gamma_{\lambda_2=\frac{1}{2}}/dq^2-d\Gamma_{\lambda_2=-\frac{1}{2}}/dq^2}{d\Gamma_{\lambda_2=\frac{1}{2}}/dq^2+d\Gamma_{\lambda_2=-\frac{1}{2}}/dq^2},
\end{eqnarray}
where,
\begin{eqnarray}
	\notag
	\frac{d\Gamma _{\lambda_2=\pm\frac{1}{2}}}{dq^2} &&= \frac{G_F^2V_{cb}^2q^2}{384\pi ^3m_{\Lambda _b[\Xi _b]}^2}\frac{\sqrt{Q_ + Q_ - }}{2m_{\Lambda _b[\Xi _b]}}\left(1 - \frac{m_\ell^2}{q^2} \right)^2\\
	\notag
	&&\times \left[ \frac{4m_\ell^2}{3q^2}\left(|H_{\pm\frac{1}{2},1}|^2 + |H_{\pm\frac{1}{2},0}|^2+3|H_{\pm\frac{1}{2},t}|^2\right) \right.\\
	&&\left. { + \frac{8}{3}\left(|H_{\pm\frac{1}{2},1}|^2 + |H_{\pm\frac{1}{2},0}|^2\right)} \right].
\end{eqnarray}
The $q^2$ dependence of $A_{FB}$ and $\alpha$ for the decay process $\Lambda_b[\Xi_b]\to\Lambda^*_c[\Xi^*_c]\ell\bar{\nu}_\ell$ are shown in Figs.~\ref{DWLX} (c), (d), (g) and (h). one can see that the leptonic forward-backward asymmetry $A_{FB}$ of these decay modes is all going to 0 near the zero recoil point $q^2=(m_{\Lambda_b[\Xi_b]}-m_{\Lambda^*_c[\Xi^*_c]})^2$ for different leptonic decay channels. However, the behaviors of different leptonic decay channels are clearly different near the maximum recoil point $q^2=m_\ell^2$. The central value of $A_{FB}$ is going to $-0.50$ and $-0.38$ for $e(\mu)$ and $\tau$ decay channels, respectively. For the asymmetric parameter $\alpha$, its central value increases from $-0.92$ to $0$ and $-0.93$ to $0$ as $q^2$ increases from $m_\ell^2$ to
$(m_{\Lambda_b[\Xi_b]}-m_{\Lambda^*_c[\Xi^*_c]})^2$ for the decay processes $\Lambda_b[\Xi_b]\to\Lambda^*_c[\Xi^*_c]\ell\bar{\nu}_\ell$ ($\ell=e$ or $\mu$) and $\Lambda_b[\Xi_b]\to\Lambda^*_c[\Xi^*_c]\tau\bar{\nu}_\tau$, respectively. By separately integrating the numerators and denominators in Eq.~(\ref{eq:66}) over the physical $q^2$
region, the mean values of $A_{FB}$ and $\alpha$ are obtained. The numerical results for these asymmetry parameters are listed in Table~\ref{AFBalpha}. Future high energy physics experiments can measure these parameters and compare them with our predictions, which would help deepen our understanding of the decay properties of singly heavy baryons and provide an opportunity to probe new physics beyond the Standard Model.
\begin{table}[htbp]
	\begin{ruledtabular}\caption{The mean values of leptonic forward-backward asymmetry $\langle A_{FB} \rangle$ and asymmetry parameter $\langle\alpha\rangle$ for semileptonic decays $\Lambda_b\to\Lambda^*_c\ell^-\bar{\nu}_\ell$ and $\Xi_b\to\Xi^*_c\ell^-\bar{\nu}_\ell$. }
		\label{AFBalpha}
		\renewcommand\arraystretch{1.3}
		\begin{tabular}{c c c}
			Decay channel&$\langle A_{FB}\rangle$&$\langle \alpha\rangle$ \\ \hline
			$\Lambda_b\to\Lambda^*_ce^-\bar{\nu}_e$&$0.167^{+0.000}_{-0.002}$&$-0.949^{+0.005}_{-0.001}$ \\
			$\Lambda_b\to\Lambda^*_c\mu^-\bar{\nu}_\mu$&$0.154^{+0.001}_{-0.001}$&$-0.948^{+0.004}_{-0.003}$ \\
			$\Lambda_b\to\Lambda^*_c\tau^-\bar{\nu}_\tau$&$-0.118^{+0.001}_{-0.002}$&$-0.899^{+0.017}_{-0.016}$ \\
			$\Xi_b\to\Xi^*_ce^-\bar{\nu}_e$&$0.185^{+0.010}_{-0.008}$&$-0.966^{+0.014}_{-0.013}$ \\
			$\Xi_b\to\Xi^*_c\mu^-\bar{\nu}_\mu$&$0.174^{+0.009}_{-0.009}$&$-0.966^{+0.014}_{-0.013}$ \\
			$\Xi_b\to\Xi^*_c\tau^-\bar{\nu}_\tau$&$-0.092^{+0.013}_{-0.011}$&$-0.927^{+0.026}_{-0.024}$ \\
		\end{tabular}
	\end{ruledtabular}
\end{table}

With the Eqs.~(\ref{eq:48}) and (\ref{eq:51}), the two-body nonleptonic decay widths of $\Lambda_b\to\Lambda^*_cP[V]$ and $\Xi_b\to\Xi^*_cP[V]$ can be obtained. The numerical results
of these two-body decay widths and branching ratios are explicitly shown in Table~\ref{DWBRNO}. According to the PDG, the product of the branching ratios for the decay $\Lambda^0_b\to\Lambda^{*+}_c(2595)\pi^-$ and the subsequent decay $\Lambda^{*+}_c(2595)\to\Lambda^+_c\pi^+\pi^-$ is $(3.4\pm1.4)\times10^{-4}$, while the branching ratio of three-body decay channel $\Lambda^{*+}_c(2595)\to\Lambda^+_c\pi^+\pi^-$ is $(40\pm14)\%$~\cite{ParticleDataGroup:2026mpi}. From these values, we obtain the branching ratio for $\Lambda^0_b\to\Lambda^{*+}_c(2595)\pi^-$ to be approximately $(0.85\pm0.46)\times10^{-3}$, which is compatible with our result.
\begin{table}[htbp]
	\renewcommand\arraystretch{1.3}
	\begin{ruledtabular}
		\caption{Decay widths (in $10^{-16}$ GeV) and Branching ratios (in $10^{-3}$) of $\Lambda_{b}$ and $\Xi_{b}$ nonleptonic decays.}
		\begin{tabular}{c >{\centering}p{6.5em}| c c }\label{DWBRNO}
			&Decay channels & Decay widths & Branching ratios\\ \hline
			&$\Lambda_b\to\Lambda^*_c\pi$ &$4.784^{+0.598}_{-0.563}$ &$1.067^{+0.133}_{-0.125}$ \\
			&$\Lambda_b\to\Lambda^*_cK$ &$0.370^{+0.046}_{-0.044}$ &$0.0825^{+0.0103}_{-0.0100}$ \\
			&$\Lambda_b\to\Lambda^*_c\rho$ &$12.734^{+1.583}_{-1.489}$ &$2.840^{+0.353}_{-0.332}$ \\
			&$\Lambda_b\to\Lambda^*_cK^*$ &$0.674^{+0.084}_{-0.079}$ &$0.150^{+0.019}_{-0.018}$ \\
			&$\Xi_b\to\Xi^*_c\pi$ &$3.048^{+0.600}_{-0.553}$ &$0.684^{+0.135}_{-0.124}$ \\
			&$\Xi_b\to\Xi^*_cK$ &$0.236^{+0.047}_{-0.043}$ &$0.0530^{+0.0104}_{-0.0096}$ \\
			&$\Xi_b\to\Xi^*_c\rho$ &$8.255^{+1.558}_{-1.446}$ &$1.852^{+0.350}_{-0.324}$ \\
			&$\Xi_b\to\Xi^*_cK^*$ &$0.439^{+0.082}_{-0.076}$ &$0.0986^{+0.0184}_{-0.0171}$ \\
		\end{tabular}
	\end{ruledtabular}
\end{table}
\section{Conclusion}\label{sec6}
In this work, we first reanalyze the mass and pole residue of $\Lambda$-type singly heavy baryons with spin-parity $J^P=1/2^{\pm}$ using two-point QCD sum rules. Then, we study the electroweak transition form factors of $\Lambda_b\to\Lambda_c^*(2595)$ and $\Xi_b\to\Xi_c^*(2790)$ using three-point QCD sum rules. At the hadron side, we take into account all possible couplings of the interpolating currents to hadronic states, and at the QCD side, we include the perturbative part and vacuum condensate contributions up to dimension-8. The form factors are obtained in the space-like region and extrapolated to the time-like region using the $z$-series expansion approach. Using the obtained form factors, we calculate the decay widths and branching ratios for the semileptonic decays $\Lambda_b\to\Lambda_c^*\ell\bar{\nu}_\ell$ and $\Xi_b\to\Xi_c^* \ell\bar{\nu}_\ell$ ($\ell=e,\mu,\tau$). Our estimates for $\Lambda_b\to\Lambda_c^* \ell\bar{\nu}_\ell$ are comparable with the experimental data from the PDG and other theoretical predictions, including those from LCSR, CCQM and LFQM. For $\Xi_b\to\Xi_c^* \ell\bar{\nu}_\ell$, our findings can provide valuable guidance for future experimental measurements. We also analyze several observables that are sensitive to the underlying dynamics, including the lepton universality ratios $R_{\Lambda_c^*}$ and $R_{\Xi_c^*}$, the forward-backward asymmetry $A_{FB}$, and the asymmetry parameter $\alpha$. The predictions for these observables may help to test the Standard Model and to probe possible new physics effects. Furthermore, we study the two-body nonleptonic decays $\Lambda_b\to\Lambda_c^* P[V]$ and $\Xi_b\to\Xi_c^* P[V]$ using the factorization approach. The branching ratio for $\Lambda_b^0\to\Lambda_c^{*+}(2595)\pi^-$ extracted from the PDG data is approximately $(0.85\pm 0.46)\times 10^{-3}$, which is compatible with our prediction. In summary, our results provide a comprehensive theoretical analysis of the weak decays of $\Lambda_b$ and $\Xi_b$ into the negative-parity baryons $\Lambda_c^*(2595)$ 
and $\Xi_c^*(2790)$. These results are expected to be useful for future experimental studies at LHCb.

\section*{Acknowledgments}
This work is supported by National Natural Science Foundation of China under the Grant Nos. 12335001, 12175068, as well as supported, in part, by National Key Research and Development Program under Grant No. 2024 YFA1610503 and Natural Science Foundation of HeBei Province under the Grant No. A2024502002.

\bibliographystyle{apsrev4-1}
\bibliography{ref}

\end{document}